\documentclass[10pt,journal,compsoc]{IEEEtran}
\ifCLASSOPTIONcompsoc
  \usepackage[nocompress]{cite}
\else
  \usepackage{cite}
\fi
\ifCLASSINFOpdf
\else
\fi
\usepackage{amsmath}
\usepackage{algorithmic}

\usepackage{array}

\ifCLASSOPTIONcompsoc
  \usepackage[caption=false,font=footnotesize,labelfont=sf,textfont=sf]{subfig}
\else
  \usepackage[caption=false,font=footnotesize]{subfig}
\fi

\usepackage{stfloats}
\usepackage{url}

\usepackage{amsmath}
\usepackage{graphicx}
\usepackage[colorinlistoftodos]{todonotes}
\usepackage[colorlinks=true, allcolors=blue]{hyperref}

\usepackage{comment}
 
\usepackage{color}
\definecolor{Orange}{rgb}{1,0.5,0}
\definecolor{Red}{rgb}{1,0,0}
\definecolor{Blue}{rgb}{0,0,1}

\usepackage{bm}
\usepackage{amsfonts} 
\usepackage{amssymb} 
\usepackage{stmaryrd} 

\usepackage{mathtools}

\usepackage{booktabs}
\usepackage{longtable}
\usepackage{array}
\usepackage{multirow}

\usepackage[linesnumbered,ruled,vlined]{algorithm2e}

\usepackage{enumitem}
\setlist[itemize]{leftmargin=*}

\begin{document}
%
\title{Optimizing Area Under the Curve Measures via Matrix Factorization for Predicting Drug-Target Interaction with Multiple Similarities}
%
%
%
%

\author{Bin~Liu and Grigorios~Tsoumakas
\IEEEcompsocitemizethanks{\IEEEcompsocthanksitem B. Liu and G. Tsoumakas are with the School of Informatics, Aristotle University of Thessaloniki, Thessaloniki 54124, Greece. \protect\\  E-mail: \{binliu, greg\}@csd.auth.gr}. 

\thanks{© 20xx IEEE. Personal use of this material is permitted. Permission from IEEE must be obtained for all other uses, including reprinting/republishing this material for advertising or promotional purposes, collecting new collected works for resale or redistribution to servers or lists, or reuse of any copyrighted component of this work in other works. \\ 
This work has been submitted to the IEEE for possible publication. Copyright may be transferred without notice, after which this version may no longer be accessible.} 

}

%
%

\markboth{IEEE/ACM Transactions on Computational Biology and Bioinformatics}%
{Shell \MakeLowercase{\textit{et al.}}: Bare Demo of IEEEtran.cls for Computer Society Journals}
%



\IEEEtitleabstractindextext{%
\begin{abstract}
In drug discovery, identifying drug-target interactions (DTIs) via experimental approaches is a tedious and expensive procedure. 
Computational methods efficiently predict DTIs and recommend a small part of potential interacting pairs for further experimental confirmation, accelerating the drug discovery process. 
Although it has been shown that fusing heterogeneous drug and target similarities can improve the prediction ability, the existing similarity combination methods ignore the interaction consistency for neighbour entities which is more crucial for the DTI prediction model.
Furthermore, area under the precision-recall curve (AUPR) that emphasizes the accuracy of top-ranked pairs and area under the receiver operating characteristic curve (AUC) that heavily punishes the existence of low ranked interacting pairs are two widely used evaluation metrics in DTI prediction. However, the two metrics are seldom considered as losses within existing DTI prediction methods. 
This paper first proposes two matrix factorization (MF) methods that optimize AUPR and AUC using convex surrogate losses respectively, and then develops an ensemble MF approach takes advantage of the two area under the curve metrics by combining the two single metric based MF models. Both three proposed approaches incorporate a novel local interaction consistency aware similarity interaction method to generate fused drug and target similarities that preserve vital information from the more reliable view.
Experimental results over five datasets under different prediction settings show that the proposed methods outperform various competitors in terms of the metric(s) they optimize. In addition, the validation on the top ranked novel predictions confirms the ability of our methods to discover potential new DTIs. 


\end{abstract}

\begin{IEEEkeywords}
Drug-target interaction prediction, matrix factorization, AUPR optimization, AUC optimization, local interaction consistency.
\end{IEEEkeywords}}

\maketitle

\IEEEdisplaynontitleabstractindextext

%
\IEEEpeerreviewmaketitle

\section{Introduction}
\IEEEPARstart{I}{dentifying} drug-target interactions (DTIs) is a key step of the drug discovery process~\cite{Bagherian2021MachinePaper}. 
However, the verification of DTIs via \textit{in vitro} experiments is still costly and time-consuming. Computational approaches to predicting the affinities of drug-target pairs, reduce the number of candidate interactions for further experimental validation, expediting the discovery of new drugs. 

There are two kinds of traditional computational methods for predicting DTIs. The first one is ligand-based methods, which compare the new query ligand and a set of known ligands with target proteins but perform poorly when the number of known ligands is insufficient~\cite{Jacob2008Protein-ligandApproach}. The other one is docking-based methods that require 3D structures of protein targets to conduct the docking simulation, hence they cannot handle proteins whose structure information is unavailable, e.g., G-protein coupled receptors (GPCRs)~\cite{Opella2013StructureSpectroscopy}.

In the last decade, there has been an increasing interest in predicting DTIs via chemogenomic computational methods due to their promising performance and capability to incorporate widely abundant biological data for both drugs and targets~\cite{Ezzat2018ComputationalSurvey}.
The prevalent chemogenomic computational methods usually rely on machine learning techniques, such as matrix factorization (MF)~\cite{Ezzat2017Drug-targetFactorization}, kernel machines~\cite{Airola2018FastTrick}, network mining~\cite{Olayan2018DDR:Approaches}, and deep learning~\cite{Xuan2021IntegratingPrediction}.

Nowadays, with the development of biological databases, the exploitation of diverse similarities retrieved from heterogeneous data sources leads to improvements in the accuracy of DTI prediction~\cite{Nascimento2016APrediction, Olayan2018DDR:Approaches, Wan2019NeoDTI:Interactions}.
\textit{Similarity integration}, also known as \textit{similarity fusion}, is an effective strategy to handle multiple similarities. It is usually adopted in DTI prediction models as a pre-processing step, generating a combined similarity matrix that preserves crucial information across the different data aspects. 

In addition, most DTI prediction models infer new DTIs based on the \textit{guilt-by-association} assumption~\cite{Olayan2018DDR:Approaches}, i.e. similar drugs tend to interact with the same targets and vice versa. Therefore, the combined similarity should maximize the \textit{local interaction consistency}: proximate drugs or targets in the combined similarity space should have similar interactions. 
However, existing similarity integration approaches either totally overlook the interaction information~\cite{Olayan2018DDR:Approaches} or rely on \textit{global} interaction consistency expressed as the alignment between the combined and ideal similarity derived from the interaction matrix~\cite{Nascimento2016APrediction,Ding2020IdentificationFusion}.

Furthermore, area under the precision-recall curve (AUPR) and area under the receiver operating characteristic curve (AUC) are two crucial threshold-independent evaluation metrics in DTI prediction~\cite{Schrynemackers2013OnNetworks}, where the threshold for whether to biologically validate a drug-target pair or not is decided by medicinal experts.
AUPR severely penalizes highly ranked non-interacting pairs~\cite{Davis2006TheCurves}, encouraging the top part of the prediction list to contain more interacting pairs, which is crucial for the efficiency of further {\it in vitro} experiments. 
AUC on the other hand, emphasizes minimizing the number of lower ranked interacting pairs, which is also important for DTI prediction, because the low ranked pairs are usually excluded from experimental verification, leading to undiscovered drugs and drug side-effects~\cite{Pliakos2020Drug-targetReconstruction}. 

Due to the importance of AUPR and AUC, optimizing these two metrics is expected to lead to better DTI prediction results.  
Although a few methods directly optimizing AUC (or ranking loss) do exist, they only work for limited settings, such as predicting interactions between candidate drugs and a specific target~\cite{Golkov2020DeepFunctions} or a fixed set of targets~\cite{Peska2017Drug-targetApproach}. However, important practical applications, such as drug repositioning, require identifying the interactions of existing drugs with novel targets~\cite{Chen2018MachinePrediction}.
In addition, to the best of our knowledge, there is no DTI prediction method for optimizing AUPR. 

MF, initially designed for recommendation systems, has been successfully applied to DTI prediction. Nevertheless, the square loss~\cite{Zheng2013CollaborativeInteractions} and logistic loss~\cite{Liu2016NeighborhoodPrediction} are mostly used in existing MF methods for DTI prediction.
There are two MF approaches for the recommendation task that optimize user-based area under the curve metrics~\cite{Shi2012TFMAP:Recommendation,dhanjal2015auc}. However, they cannot be applied to DTI prediction, because they fail to handle the cold start issue present in DTI prediction~\cite{Pahikkala2015TowardPredictions} and their user-based metrics are unable to fully capture the information of the interaction matrix.


In this paper, we first present a novel similarity integration approach based on Local Interaction Consistency (LIC). Our approach linearly combines diverse similarities using a weighting strategy that not only retains crucial topology across diverse data aspects, but also explicitly emphasizes the similarity possessing the higher interaction consistency for proximate drugs and targets. 
Then, we derive two surrogate AUPR and AUC losses based on drug-target pairs, upon which two MF models for DTI prediction, namely MFAUPR and MFAUC, that optimize AUPR and AUC respectively, are proposed. Finally, we propose an ensemble MF approach, MF2A, that integrates MFAUPR and MFAUC. All three MF based models adopt LIC to aggregate multiple input similarities obtained from various sources, and determine the optimal decay coefficient within a faster selection process to infer reliable latent features of novel drugs and targets involved in the test pairs.
Extensive experiments under four different prediction settings demonstrate the effectiveness of the proposed similarity integration and DTI prediction methods. Moreover, the practical ability of MF2A to predict novel DTIs is also verified by finding supportive evidence from online biological databases.

The remainder of this paper is structured as follows. In Section 2, related work is briefly reviewed. In Section 3, the problem formulation is introduced. In Section 4, the LIC and the three MF models are presented. The experimental results are discussed in Section 5. Finally, the conclusions are given in Section 6.





\section{Related Work}
In this section, we briefly review representative DTI prediction methods, and summarize the losses they optimize. The prevalent similarity integration methods employed in the DTI prediction task are discussed as well. 

Neighbourhood based methods utilize the information of neighbours to infer the unseen interactions.
A simple solution is to compute a weighted average of the interaction profiles of neighbour drugs or targets as the predictions~\cite{Yamanishi2008PredictionSpaces}. In~\cite{Shi2015PredictingClustering}, a set of super-targets are constructed by clustering similar targets to recover missing interactions, and a neighbourhood based multi-label classifier, namely MLkNN, is employed to perform prediction based on maximum a posteriori (MAP) rule. WkNNIR performs interaction recovery in both drug and target sides and exploits the local imbalance to distinguish the importance of drug and target spaces within its prediction function~\cite{Liu2021Drug-TargetRecovery}.

By treating drug and target similarities as kernels, kernel based classifiers could be applied to DTI prediction. RLS-avg~\cite{Mei2013Drug-targetNeighbors} trains two kernel regularized least squares classifiers (RLS) for drug and target respectively, and combines the outputs of the two models as its final prediction. RLS-Kron~\cite{vanLaarhoven2013PredictingProfile} trains one Kronecker kernel based RLS, but the utilization of the Kronecker product of drug and target kernels leads to cubic computational complexity, which constrains its scalability. KSVM~\cite{Airola2018FastTrick} is a Kronecker kernel based support vector machine (SVM) model optimizing the square hinge loss, which employs a generalized vec trick to accelerate the computation on the Kronecker product, yielding linear computational complexity. To handle diverse drug and target similarities (kernels), multiple kernel learning, which aims to obtain an optimal combination of kernels, is a prevalent strategy in kernel based methods. KRLSM~\cite{Nascimento2016APrediction} incorporates kernel integration and RLS-Kron training in a uniform framework. DRLSM~\cite{Ding2020IdentificationFusion} obtains optimal combined drug and target kernels that are aligned with the ideal kernel derived from the interaction matrix following the Hilbert–Schmidt Independence Criterion (HSIC), and then trains a dual Laplacian regularized RLS that incorporates the two optimal kernels to perform predictions.   

MF DTI prediction methods find latent drug and target feature matrices, whose product is able to approximate the interaction matrix. Most MF methods employ square loss to assess the difference between the estimated and ground truth interaction matrices. 
GRMF~\cite{Ezzat2017Drug-targetFactorization} performs graph regularization on latent features to learn a manifold for label propagation in drug and target spaces. Neighbourhood based interaction replenishment pre-processing is used by GRMF to address the influence of possible missing interactions. MSCMF~\cite{Zheng2013CollaborativeInteractions} ensures that similar drugs (targets) tend to be close in latent feature space by adding the low-rank decomposition regularization on the fused similarity matrix, which linearly combines multiple types of similarity. 
MLRE~\cite{Li2019DrugEmbedding} is a multi-view low rank embedding method, which learns view-wise drug and target low-rank features in a common embedding space, and encourages the similarity of features across all views by co-regularized spectral clustering.
NRLMF~\cite{Liu2016NeighborhoodPrediction} combines logistic matrix factorization with neighbourhood regularization in a unified framework, which essentially optimizes the logistic loss. Peska et al.~\cite{Peska2017Drug-targetApproach} formulate DTI prediction as a per-drug ranking problem, i.e. an interacting target should be ranked higher than a non-interacting one for each drug, and propose a content alignment regularized Bayesian personalized ranking MF method (BRDTI), which utilizes the drug-wise ranking loss in its objective function. 

A DTI dataset could be considered as a heterogeneous network, where drugs and targets are two kinds of nodes, and edges are represented by adjacency matrices containing DTIs, as well as drug and target similarities.
Network diffusion techniques, such as network inference ~\cite{Wang2013DrugInference} and random walk~\cite{Chen2012Drug-targetNetwork}, are applied on the DTI heterogeneous network to predict new DTIs. 
Apart from inferring unknown interactions on the DTI heterogeneous network directly, another solution is to learn low-dimensional drug and target embeddings that capture the topological structure of the network to facilitate the DTI prediction. 
In~\cite{Olayan2018DDR:Approaches,Chu2021DTI-CDF:Features,Thafar2020DTiGEMS+:Techniques}, path category based features, reflecting the statistics of six types of short paths going through the corresponding drug or target, are extracted.
Luo et al.~\cite{Luo2017AInformation} proposed a DTI prediction pipeline called DTINet, which learns drug and target embeddings upon a heterogeneous network that incorporates diverse drug and target-related information via compact feature learning, and optimizes a projection matrix to map the drug and target embedding space to the output interaction space. NeoDTI~\cite{Wan2019NeoDTI:Interactions} generates the drug and target features via neighbourhood information aggregation and single-layer neural network based nonlinear embedding transformation. 

With the increasing popularity of deep learning, graph neural models are adopted for network based DTI prediction. GANDTI~\cite{Sun2020GraphInteractions} utilizes an adversarially regularized graph autoencoder to learn drug and target representations that follow the Gaussian distribution. DTIP~\cite{Xuan2021IntegratingPrediction} leverages a feature-level attention based autoencoder and the bidirectional gated recurrent unit to derive the drug and target representations from a multiplex heterogeneous DTI network, and employs a convolutional neural network to predict unknown DTIs.

Some methods treat interaction prediction as a classification task, building binary or multi-label classifiers over standard or derived feature sets. EBiCTR~\cite{Pliakos2020Drug-targetReconstruction} is an ensemble of bi-clustering trees trained on the reconstructed output space and dyadic (drug and target) feature space. Three tree-based multi-label classifiers that incorporate various label partition strategies to effectively capture the correlations among drugs and targets are proposed for DTI prediction in ~\cite{Pliakos2021PredictingPartitioning}. 

Regarding feature generation, molecular descriptors representing the characteristics of the chemical structure of drugs and protein descriptors illustrating the structural and physicochemical properties of the genomic sequence of targets are conventional artificially designed representations~\cite{Sachdev2019APrediction}. Recently, deep learning models are employed to learn more robust high-level representations from raw chemical structures of drugs and amino acid sequences of targets~\cite{Gao2018InterpretableRepresentation, Yu2021KenDTI:Prediction}. 

Furthermore, we briefly summarize the losses that DTI prediction methods optimize. The square loss is the most commonly used by the majority of kernel~\cite{Mei2013Drug-targetNeighbors, vanLaarhoven2013PredictingProfile,Nascimento2016APrediction,Ding2020IdentificationFusion}, MF~\cite{Ezzat2017Drug-targetFactorization,Zheng2013CollaborativeInteractions,Li2019DrugEmbedding}, network~\cite{Olayan2018DDR:Approaches,Luo2017AInformation,Wan2019NeoDTI:Interactions,Sun2020GraphInteractions} and standard classification~\cite{Pliakos2020Drug-targetReconstruction,Pliakos2021PredictingPartitioning} methods. In addition, the logistic (cross entropy) loss is employed by probability estimation methods~\cite{Liu2016NeighborhoodPrediction} as well as deep learning models~\cite{Xuan2021IntegratingPrediction,Yu2021KenDTI:Prediction}, and hinge loss is optimized by SVM  approaches~\cite{Airola2018FastTrick}.
However, these frequently-used losses are typically biased to non-interacting pairs that predominate the dataset, leading to failures in distinguishing interacting ones.
The ranking loss or AUC based loss is a substitution, which is insusceptible to the frequency of interacting pairs via punishing inversely ranked interacting and non-interacting pairs ~\cite{Peska2017Drug-targetApproach,Golkov2020DeepFunctions}. 

Lastly, we review similarity integration strategies used for DTI prediction. Similarity integration methods are model-agnostic, i.e. they could work with any prediction model that handles a single type of similarity, and enable to improve the computational efficiency of the prediction model via shrinking the dimension of the input space. The most intuitive solution, AVE~\cite{Nascimento2016APrediction}, is to average multiple similarity matrices. Similarity Network Fusion (SNF)~\cite{Olayan2018DDR:Approaches} aggregates various similarities via a non-linear fusion process. However, these two methods do not take interaction information into account. On the other hand, Kernel Alignment (KA)~\cite{Nascimento2016APrediction} explicitly exploits the interaction information by assigning a higher weight to a similarity matrix that is more aligned to the ideal similarity derived from the interaction matrix. HSIC~\cite{Ding2020IdentificationFusion} learns an optimally combined similarity that maximizes the dependence of the ideal interaction similarity via a HSIC-based kernel fusion framework. Both KA and HSIC utilize the similarity alignment or dependence on the whole space, without any emphasis on interaction consistency for neighbour drugs (targets). 

\section{Problem Formulation}
We introduce the formulation of the DTI prediction problem, where input information are multiple drug and target similarities. 
Let $D=\{d_i\}_{i=1}^{n_d}$ be a set of drugs and $T=\{t_i\}_{i=1}^{n_t}$ be a set of targets, where $n_d$ and $n_t$ are the number of drugs and targets, respectively. Let $\{\bm{S}^{d,h}\}_{h=1}^{m_d}$ be a set of drug similarity matrices, where $\bm{S}^{d,h} \in \mathbb{R}^{n_d \times n_d}$ and $m_d$ is the number of drug similarity matrices. Similarly, let $\{\bm{S}^{t,h}\}_{h=1}^{m_t}$ be a set of target similarity matrices, where $\bm{S}^{t,h} \in \mathbb{R}^{n_t \times n_t}$ and $m_t$ is the number of target similarity matrices. The interactions between $D$ and $T$ are represented as a binary matrix $\bm{Y} \in \{0,1\}^{n_d \times n_t}$, where $Y_{ij}=1$ denotes $d_i$ and $t_j$ interact with each other, and $Y_{ij} = 0$ otherwise. The DTI training set for $D$ and $T$ consists of $\{\bm{S}^{d,h}\}_{h=1}^{m_d}$, $\{\bm{S}^{t,h}\}_{h=1}^{m_t}$ and $\bm{Y}$.

Let ($d_x$,$t_z$) be a test pair, $\{\bm{\bar{s}}^{d,h}_x\}^{m_d}_{h=1}$ be a set of $n_d$-dimensional vectors storing the similarities between $d_x$ and $D$, and $\{\bm{\bar{s}}^{t,h}_z\}^{m_d}_{h=1}$ be a set of $n_t$-dimensional vectors storing the similarities between $t_z$ and $T$. A DTI prediction model predicts a real-valued score $\hat{Y}_{xz}$ indicating the confidence of the affinity between $d_x$ and $t_z$. 
There are four prediction settings according to whether the drug and target involved in the test pair are included in the training set or not~\cite{Pahikkala2015TowardPredictions}: 
\begin{itemize}
    \item S1: predict the interaction between $d_{x} \in D$ and $t_z \in T$
    \item S2: predict the interaction between $d_{x} \notin D$ and $t_z \in T$
    \item S3: predict the interaction between $d_x \in D$ and $t_z \notin T$
    \item S4: predict the interaction between $d_x \notin D$ and $t_z \notin T$
\end{itemize}
where $d_x \notin D$ ($t_z \notin T$) is the new drug (target).

In this paper, given a matrix $\bm{A}$, $\bm{A}_{i}$ denotes its $i$-th row, and $A_{ij}$ represents its element in $i$-th row and $j$-th column. 

\section{Methods}
\begin{figure*}[th]
\centering
\includegraphics[width=0.7\textwidth]{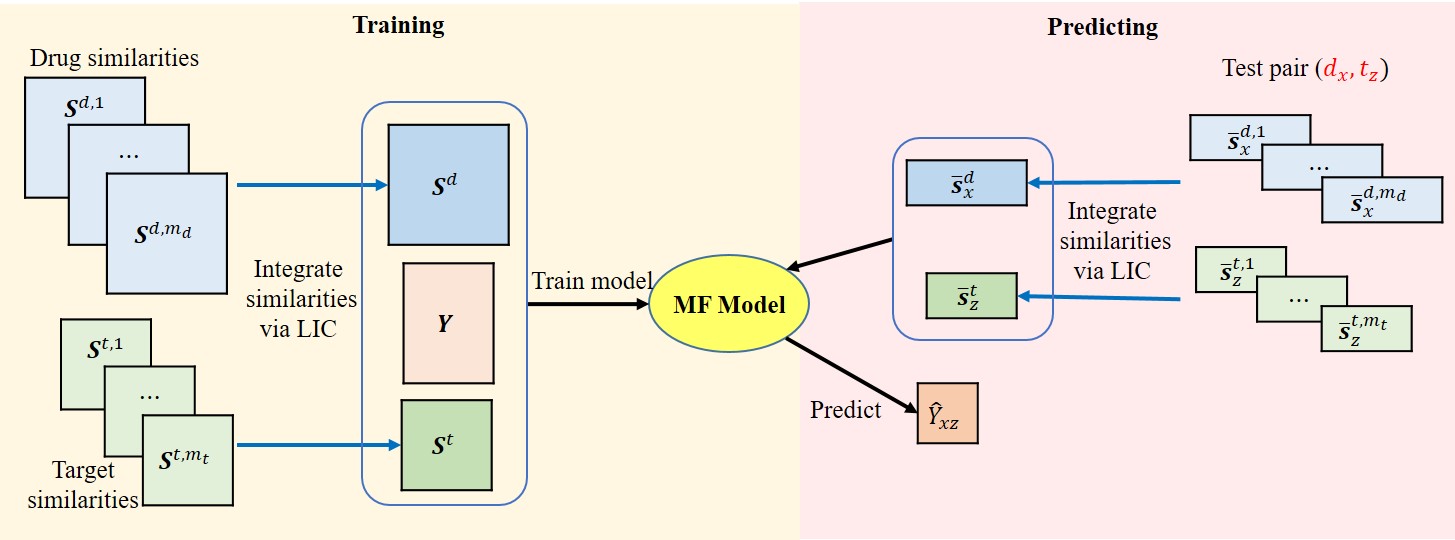}
\caption{The workflow of proposed methods, where blue lines represent similarity integration, black lines denotes model training and test procedures, and MF model could be MFAUPR, MFAUC or MF2A.} 
\label{fig:Workflow}
\end{figure*}

We propose three MF based DTI prediction methods, namely MFAUPR, MFAUC and MF2A, that optimize AUPR, AUC and both these two metrics, respectively. All three methods follow the same workflow, which is shown in Fig. \ref{fig:Workflow}. In both training and predicting phases, the similarity integration is firstly conducted to combine multiple similarities into a fused one, which is then used to train the MF model or obtain the prediction of the test pair. 

In this section, we first present a similarity integration approach that emphasizes similarities having more reliability based on the local interaction consistency. Then, the general formulation of MF model for DTI prediction is introduced, and three MF models are proposed. Lastly, we analyse the computational complexity of the proposed methods.

\subsection{LIC based Similarity Integration}

Multiple drug (target) similarities that capture relations between drugs (targets) in various aspects could be linearly combined into a new one as follows:   
\begin{equation}
    \bm{S}^{\alpha} = \sum_{h=1}^{m_\alpha} w^{\alpha}_h \bm{S}^{\alpha,h}, \quad \alpha \in \{d,t\}
\end{equation}
where $w^{\alpha}_h$ is the weight of $\bm{S}^{\alpha,h}$, and the more important similarities are usually endowed with a larger weight. 
In this part, we propose a linear similarity integration method, termed LIC, which determines similarity weights according to the local interaction consistency.


\textit{Local imbalance}, which was firstly introduced to assess the class discrepancy within local area~\cite{liu2021multi}, has been also found to be an effective measurement to estimate the quality of drug and target similarities in DTI prediction~\cite{Liu2021Drug-TargetRecovery}. 
The local imbalance of a drug is defined as the proportion of its neighbours having different interactivity, which reflects the \textit{local interaction discrepancy} of $d_i$.
Inspired by the concept of local imbalance, we propose a new measure to evaluate the local interaction consistency of similarities. Specifically, the local interaction consistency of a drug $d_i$ for one of its interacting targets $t_j$ based on the $h$-th drug similarity is defined as:
\begin{equation}
B^{d,h}_{ij} = \frac{1}{\sum_{d_l \in {\cal N}^{k,h}_{d_i}}S^{d,h}_{il}} \sum_{d_l \in {{\cal N}^{k,h}_{d_i}}}  S^{d,h}_{il} \llbracket Y_{il} = Y_{ij} \rrbracket
\label{eq:Bdh_ij}
\end{equation}
where ${\cal N}^{k,h}_{d_i}$ denotes the $k$-nearest neighbours of $d_i$ retrieved according to $\bm{S}^{d,h}_i$ and $\llbracket \cdot \rrbracket$ is the indicator function that returns 1 if the input event is true and 0 otherwise. Compared to the local imbalance of $d_i$ defined in~\cite{Liu2021Drug-TargetRecovery}, the similarity value ($S^{d,h}_{il}$) is added in Eq.\eqref{eq:Bdh_ij} to emphasize the importance of more similar drugs. Higher $B^{d,h}_{ij}$ values reflect that $d_i$ keeps higher interaction consistency in $\bm{S}^{d,h}$, as more similar drugs of $d_i$ interact with $t_j$, which further indicates the larger reliability of $\bm{S}^{d,h}$ for $d_i$ regarding the interactivity with $t_j$. 
Averaging the local interaction consistency of all drugs with respect to their corresponding interacting targets, we arrive at the local interaction consistency of $\bm{S}^{d,h}$:
\begin{equation}
    LC^d_h = \frac{1}{|P_1|} \sum_{(i,j) \in P_1} B^{d,h}_{ij}
\label{eq:LCd}
\end{equation} 
where $P_1 =\{(i,j)|Y_{ij}=1\} $ is the set of all interacting drug-target pairs.
Based on $LC^d_h$, we define the weight of $\bm{S}^{d,h}$ as:
\begin{equation}
    w^d_h = \frac{LC^d_h}{\sum_{i=1}^{m_d} LC^d_i}
    \label{eq:wd_h}
\end{equation}
where the denominator is a normalization term ensuring the sum of all weights to be 1.
The weight of a target similarity $w^t_h$ is calculated in the same way, except for replacing $Y_{il}$ with $Y_{il}^\top$ in Eq.\eqref{eq:Bdh_ij} and substituting $d$ with $t$ in Eq.\eqref{eq:Bdh_ij}-\eqref{eq:wd_h}. 

For a new drug (target), its merged similarities to training drugs (targets) are also obtained by employing the local interaction consistency based weights:
\begin{equation}
    \bar{\bm{s}}^{\alpha}_x = \sum_{h=1}^{m_\alpha}w^{\alpha}_h \bar{\bm{s}}^{\alpha,h}_{x}, \quad \alpha \in \{d,t\}
\end{equation}

\subsection{MF for DTI Prediction}
In DTI prediction, MF based approaches decompose the interaction matrix to obtain vectorized representations of drugs and targets, upon which the unseen interactions are predicted. As the LIC is applied in the first step, integrated similarity matrices ($\bm{S}^d$ and $\bm{S}^t$) that captures the information of various similarities are used as the input of our MF models.

Formally, the MF method learns two $r$-dimensional low-rank latent features (embeddings) $\bm{U} \in \mathbb{R}^{n_d \times r}$ (for drugs) and $\bm{V} \in \mathbb{R}^{n_t \times r}$ (for targets) that approximate the training interaction matrix $\bm{Y}$ by minimizing the follow objective:
\begin{equation}
\min_{\bm{U},\bm{V}} {\cal J}={\cal L}(\hat{\bm{Y}},\bm{Y}) + {\cal R}(\bm{U},\bm{V})
\label{eq:MF_obj}
\end{equation}
${\cal L}(\hat{\bm{Y}},\bm{Y})$ is the loss function, $\hat{\bm{Y}} = \sigma(\bm{U}\bm{V}^\top) \in \mathbb{R}^{n_d \times n_t}$ is the predicted interaction matrix, where $\sigma$ is either the identity function $\sigma_1$ for standard MF~\cite{Ezzat2017Drug-targetFactorization} or the element-wise logistic function $\sigma_2$ for Logistic MF~\cite{Liu2016NeighborhoodPrediction}.
${\cal R}(\bm{U},\bm{V})$ concerns the regularization of the latent features and is defined as: 
\begin{align}
{\cal R}(\bm{U},\bm{V}) = & \frac{\lambda_r}{2}(||\bm{U}||_{F}^2 +||\bm{V}||_{F}^2) \nonumber\\
& + \frac{\lambda_d}{2}\text{tr}(\bm{U}^\top \bm{L}^d \bm{U}) + \frac{\lambda_t}{2}\text{tr}(\bm{V}^\top \bm{L}^t\bm{V}) 
\label{eq:RUV}
\end{align}
where $\lambda_r$, $\lambda_d$, $\lambda_t$ are regularization coefficients, while $\bm{L}^d$ and $\bm{L}^t$ are the graph Laplacians of the sparse drug and target similarity matrices, respectively. Following the sparsification process in~\cite{Liu2016NeighborhoodPrediction}, we obtain the sparse drug similarity matrix $\hat{\bm{S}}^d \in \mathbb{R}^{n_d \times n_d}$ by retaining the $k$-largest similarities in each row of $\bm{S}^{d}$, i.e. $\hat{S}^d_{ij}=S^d_{ij}$ if $d_j \in {\cal N}^k_{d_i}$, where ${\cal N}^k_{d_i}$ is $k$ nearest neighbors (most similar drugs) of $d_i$ based on $\bm{S}^d_{i}$, and $\hat{S}^d_{ij}=0$ otherwise.
The Laplacian matrix for $\hat{\bm{S}}^d$ is defined as  $\bm{L}^d=\bm{\Lambda}^d-\hat{\bm{S}}^d+\Tilde{\bm{\Lambda}}^d-(\hat{\bm{S}}^d)^\top$, where $\bm{\Lambda}^d$ and $\Tilde{\bm{\Lambda}}^d$ are two diagonal matrices with diagonal elements $\bm{\Lambda}^d_{ii}=\sum_{j=1}^m \hat{S}^d_{ij}$ and $\Tilde{\bm{\Lambda}}^d_{jj}=\sum_{i=1}^m \hat{S}^d_{ij}$, respectively.
The sparse target similarity matrix $\hat{\bm{S}}^t$ and its graph Laplacian $\bm{L}^t$ are computed in the same way.
The first term in ${\cal R}(\bm{U},\bm{V})$ applies Tikhonov regularization that prevents latent features from overfitting to the training data, while the last two terms apply graph regularization, which guarantees their local invariance~\cite{Liu2018RegularizedSurvey}, i.e. similar drugs (targets) are likely to have similar latent features. 

Given a test drug-target pair ($d_x,t_z$), the latent features of $d_x$ and $t_z$, denoted as $\bm{U}_x \in \mathbb{R}^{r}$ and $\bm{V}_z \in \mathbb{R}^{r}$ respectively, are obtained firstly. If $d_x$ ($t_z$) is a training drug (target), its latent features are directly extracted from $\bm{U}$ ($\bm{V}$), e.g. $\bm{U}_x$ is the $x$-th row of $\bm{U}$. For a new drug (target), we infer its latent features by linearly combining the embeddings of its $k$-nearest training drugs (targets) using the following equations:
\begin{align}
\bm{U}_x =  \frac{1}{\sum_{d_i \in {{\cal N}^k_{d_x}}}\bar{s}^d_{xi}} \sum_{d_i \in {{\cal N}^k_{d_x}}}\eta^{i'-1}\bar{s}^d_{xi}\bm{U}_{i}  
\label{eq:U_x} \\
\bm{V}_z =  \frac{1}{\sum_{t_j \in {{\cal N}^k_{t_z}}}\bar{s}^t_{zj}} \sum_{t_j \in {{\cal N}^k_{t_z}}}\eta^{j'-1}\bar{s}^t_{zj}\bm{V}_{j}
\label{eq:V_z}
\end{align}
where $i'$ ($j'$) is the rank of $d_i$ ($t_j$) among ${\cal N}^k_{d_x}$ (${\cal N}^k_{t_z}$), e.g. $i'$=2 if $d_i$ is the second nearest neighbor of $d_x$, $\eta \in [0,1]$ is the decay coefficient shrinking the weight of further neighbors, and $\bar{s}^d_{xi}$ ($\bar{s}^t_{zj}$) is the integrated similarity between the new drug $d_x$ (target $t_z$) and the training drug $d_i$ (target $t_j$). The decay coefficient is an important parameter for nearest neighbors aggregation~\cite{Ezzat2017Drug-targetFactorization,Liu2021Drug-TargetRecovery}, which we use in Eq. \eqref{eq:U_x} and \eqref{eq:V_z} to obtain more reliable latent features for new drugs and targets, in contrast to \cite{Liu2016NeighborhoodPrediction}.
Then, the MF model outputs its prediction with a specific instantiation of the $\sigma$:
\begin{equation}
\hat{Y}_{xz} = \left\{ 
\begin{aligned}
&\bm{U}_x\bm{V}_z^\top, \text{if } \sigma = \sigma_1 \\
&\left(1+\exp(-\bm{U}_x\bm{V}_z^\top)\right)^{-1}, \text{if } \sigma = \sigma_2 
\end{aligned}
\right. 
\end{equation}

\subsection{MFAUPR}
The computation of AUPR with linear interpolations on the precision-recall (PR) curve leads to overly-optimistic performance estimation~\cite{Davis2006TheCurves}. Therefore, we employ the uninterpolated PR curve to compute AUPR.
Given $\bm{Y}$ and its predictions $\bm{\hat{Y}}$, all predictions are firstly sorted in descending order. Let $R_{h} = (i_h, j_h)$ be the index tuple of the $h$-th largest prediction. Based on the rank list $R$, AUPR is computed as: 
\begin{equation}
\begin{aligned}
AUPR(\bm{\hat{Y}}, \bm{Y}) &= \sum_{h=1}^{n_dn_t} Prec@h * InRe@h \\ 
\end{aligned}
\label{eq:AUPR}
\end{equation}
where $Prec@h = \frac{1}{h} \sum_{i=1}^h Y_{R_{i}}$ is the precision of the first $h$ predictions, $InRe@h = Y_{R_h}/\psi({\bm{Y}})$ is the incremental recall from rank $h-1$ to $h$, and $\psi({\bm{Y}}) = \sum_i \sum_j Y_{ij}$ denotes the summation of elements in $\bm{Y}$.
The AUPR defined in Eq.\eqref{eq:AUPR} is also called \textit{Average Precision}~\cite{cambridge2009online}. 

AUPR is hard to optimize directly, because it depends on the ranks of the predictions obtained by the non-differentiable and non-smooth sorting operation. 
Histogram binning that assigns the predictions into several ordered groups to simulate the ranking process provides a differential substitution for the sorting operation~\cite{He2018HashingRank,Revaud2019LearningLoss}.
MFAUPR employs the histogram binning strategy to derive the approximate differential AUPR.
To divide histogram bins easily, MFAUPR uses the element wise logistic function $\sigma_2$ for predictions, i.e. $\hat{Y}_{ij} = (1+\exp(-\bm{U}_i\bm{V}_{j}^\top))^{-1}$, to confine all predictions in the range of $(0,1)$. 
The range of predictions is divided into $n_b$ bins (intervals), and the $h$-th bin is $\bar{b}_h = [\max(b_h-\Delta,0), \min(b_h-\Delta,1))$, where $\Delta = 1/(n_b-1)$ is the bin width and $b_h=1-(h-1)\Delta$ is the center of $\bar{b}_h$. It should be noted that $\bar{b}_1$ and $\bar{b}_M$ are half-sized bins.

The soft assignment function that returns the membership degree of a prediction to $\bar{b}_h$ is defined as:
\begin{equation}
\delta(\hat{Y}_{ij},h)= \max\left( 1-|\hat{Y}_{ij}-b_h|/\Delta,0 \right)
\label{eq:delta}
\end{equation}
where the closer $\hat{Y}_{ij}$ is to $b_h$, the higher the membership degree of $\hat{Y}_{ij}$ to $\bar{b}_h$, and vice versa.
$\delta(\hat{Y}_{ij},h)$ is a differentiable function~\cite{Revaud2019LearningLoss} and its derivative w.r.t $\hat{Y}_{ij}$ is:
\begin{equation}
\nabla_{\hat{Y}_{ij}}\delta(\hat{Y}_{ij},h) = -\frac{1}{\Delta} \text{sign}(\hat{Y}_{ij}-b_h) \llbracket |\hat{Y}_{ij}-b_h| \leq \Delta \rrbracket
\end{equation}
where $\text{sign}(\cdot)$ is the sign function that returns 1 and -1 for positive and negative input values, respectively. Furthermore, we define the matrix version of $\delta(\cdot,h)$ and its derivative as: $\delta(\bm{\hat{Y}},h) \in \mathbb{R}^{n_d \times n_t}$ with $[\delta(\bm{\hat{Y}},h)]_{ij} = \delta(\hat{Y}_{ij},h)$ and $\nabla \delta(\bm{\hat{Y}},h)\in \mathbb{R}^{n_d \times n_t}$ with $[\nabla \delta(\bm{\hat{Y}},h)]_{ij} = \nabla_{\hat{Y}_{ij}}\delta(\hat{Y}_{ij},h)$.

Based on the definition of the bin and soft assignment function, the differential approximations of the precision at $\bar{b}_h$ and the increment of recall from $\bar{b}_{h-1}$ to $\bar{b}_h$ are defined as:
\begin{equation}
Prec'@h = \frac{\sum_{i=1}^h \psi(\delta(\hat{\bm{Y}},i)\odot \bm{Y})}{\sum_{i=1}^h  \psi(\delta(\hat{\bm{Y}},i))} 
\label{eq:Prec2}
\end{equation}
\begin{equation}
InRe'@h = \frac{1}{\psi({\bm{Y}})}\psi(\delta(\hat{\bm{Y}},h)\odot \bm{Y}) 
\label{eq:InRe2}
\end{equation}
where $\odot$ is the element-wise product. 
The differential approximation of AUPR that accumulates the precision and incremental recall of every bin is:
\begin{equation}
AUPR' = \sum_{h=1}^{n_b} Prec'@h * InRe'@h
\label{eq:AUPR2}
\end{equation}

Based on Eq. \eqref{eq:Prec2}-\eqref{eq:AUPR2}, we obtain the AUPR based loss by ignoring the constant value $\psi({\bm{Y}})$:
\begin{equation}
{\cal L}_{AP} = -\sum_{h=1}^{n_b}\frac{\psi(\delta(\hat{\bm{Y}},h)\odot \bm{Y}) \sum_{i=1}^h \psi(\delta(\hat{\bm{Y}},i)\odot \bm{Y})}{\sum_{i=1}^h  \psi(\delta(\hat{\bm{Y}},i))}
\label{eq:L_AP}
\end{equation}
Minimizing ${\cal L}_{AP}$ is equivalent to maximizing $AUPR'$. 
By replacing ${\cal L}(\hat{\bm{Y}},\bm{Y})$ with ${\cal L}_{AP}$, we define the optimization problem of MFAUPR as:
\begin{equation}
\min_{\bm{U},\bm{V}} {\cal J}_{AP}={\cal L}_{AP} + {\cal R}(\bm{U},\bm{V})
\label{eq:MFAUPR_obj}
\end{equation}

Eq. \eqref{eq:MFAUPR_obj} can be solved by alternating optimization~\cite{Liu2016NeighborhoodPrediction,Ezzat2017Drug-targetFactorization}. In each iteration, $\bm{U}$ is firstly updated with fixed $\bm{V}$, and then $\bm{V}$ is updated with fixed $\bm{U}$.
For the sake of conciseness, we define the following variables: $\widetilde{\psi}_h = \psi(\delta(\hat{\bm{Y}},h))$, $\psi_h = \psi(\delta(\hat{\bm{Y}},h)\odot \bm{Y})$,  $\Psi_h= \sum_{i=1}^h \psi_i$, $\widetilde{\Psi}_h= \sum_{i=1}^h \widetilde{\psi}_i$, $\widetilde{\bm{Z}}^{h} = \nabla \delta(\bm{\hat{Y}},h) \odot \hat{\bm{Y}} \odot (\bm{1}-\hat{\bm{Y}})$, $\bm{Z}^h =\bm{Y} \odot \widetilde{\bm{Z}}^{h}$. The gradients of ${\cal L}_{AP}$ w.r.t $\bm{U}$ and $\bm{V}$ are:
\begin{align}
\nabla_{\bm{U}} {\cal L}_{AP} = & \sum_{h=1}^{n_b} \frac{\psi_h}{ (\widetilde{\Psi}_h)^2} \left( \Psi_h \sum_{i=1}^h \widetilde{\bm{Z}}^{i} \bm{V} - \widetilde{\Psi}_h \sum_{i=1}^h \bm{Z}^{i}\bm{V} \right) \nonumber\\
& - \frac{\Psi_h}{ \widetilde{\Psi}_h} \bm{Z}^{h}\bm{V}  
\end{align}
\begin{align} 
\nabla_{\bm{V}} {\cal L}_{AP} = & \sum_{h=1}^{n_b} \frac{\psi_h}{ (\widetilde{\Psi}_h)^2} \left( \Psi_h \sum_{i=1}^h \widetilde{\bm{Z}}^{{i}^{\top}}  \bm{U} - \widetilde{\Psi}_h \sum_{i=1}^h {\bm{Z}^{i}}^{\top} \bm{U} \right) \nonumber\\
& - \frac{\Psi_h}{ \widetilde{\Psi}_h} {\bm{Z}^{h}}^{\top} \bm{U} ,
\end{align}
According to Eq. \eqref{eq:RUV}, the gradients of the regularization term w.r.t $\bm{U}$ and $\bm{V}$ are:
\begin{align}
& \nabla_{\bm{U}} {\cal R}(\bm{U},\bm{V}) = \lambda_r\bm{U} + \lambda_d \bm{L}^d\bm{U}
\label{eq:deriv_RUV_U} \\
& \nabla_{\bm{V}} {\cal R}(\bm{U},\bm{V}) = \lambda_r\bm{V} + \lambda_t \bm{L}^t\bm{V}
\label{eq:deriv_RUV_U}
\end{align}
The training process of MFAUPR is shown in Algorithm \ref{alg:MFAUPR_train}, where $\theta$ is the learning rate, $k$ is the number of neighbors for similarity sparsification and  inferring latent features, and ${\cal C}$ is the set of candidate values for $\eta$.

\begin{algorithm}[t]
\caption{Training of MFAUPR}
\label{alg:MFAUPR_train}
\SetKwData{Left}{left}\SetKwData{This}{this}\SetKwData{Up}{up}
\SetKwInOut{Input}{input}\SetKwInOut{Output}{output}
\Input{$\bm{Y}$, $D$, $T$, $\bm{S}^d$, $\bm{S}^t$, $\theta$, $k$, ${\cal C}$}
\Output{$\bm{U}$, $\bm{V}$, $\eta_2$, $\eta_3$, $\eta_4$}
Compute $\bm{L}^d$ and $\bm{L}^t$ based on sparsified $\bm{S}^d$ and $\bm{S}^t$ \;
Initialize $\bm{U}$ and $\bm{V}$ randomly\;
\Repeat{convergence}{
$\nabla_{\bm{U}} {\cal J}_{AP} \leftarrow \nabla_{\bm{U}} {\cal L}_{AP} + \nabla_{\bm{U}} {\cal R}(\bm{U},\bm{V})$ \;
$\bm{U} \leftarrow \bm{U} - \theta \nabla_{\bm{U}}{\cal J}_{AP}$ \; 
$\nabla_{\bm{V}} {\cal J}_{AP} \leftarrow \nabla_{\bm{V}} {\cal L}_{AP} + \nabla_{\bm{V}} {\cal R}(\bm{U},\bm{V})$ \;
$\bm{V} \leftarrow \bm{V} - \theta \nabla_{\bm{V}}{\cal J}_{AP}$ \; 
}
Choose $\eta_2$, $\eta_3$, $\eta_4$ from ${\cal C}$ using Algorithm \ref{alg:Opti_eta}\;
\end{algorithm}


As we mentioned before, $\eta$ has crucial impact on the quality of the estimated latent features. In MFAUPR, instead of tuning $\eta$ as a hyper-parameter using grid search, which is time-consuming, the optimal $\eta$ for each prediction setting (e.g. $\eta_2$ is the optimal value for S2) is chosen based on the training set after solving Eq \eqref{eq:MFAUPR_obj}, as shown in Algorithm \ref{alg:Opti_eta}. 
Specifically, given an $\eta$ from ${\cal C}$, we compute the pseudo latent features of training drugs and targets, which along with $\bm{U}$ and $\bm{V}$ are used to compute three prediction matrices under three assumed prediction settings (S2, S3 and S4) respectively (line 8, Algorithm \ref{alg:Opti_eta}), e.g. the estimated drug, $\bm{U}'$, and target, $\bm{V}$, latent features are used to make predictions for S2.
Then, the AUPR of these prediction matrices are obtained.
The above procedure is repeated for all $\eta$ candidates. For each setting, the $\eta$ whose prediction matrix achieves the highest AUPR is the optimal one. In prediction phase, the prediction setting of the test pair is identified, and the corresponding optimal $\eta$ is used to infer the latent features of the test drug and/or target.


\begin{algorithm}[h]
\caption{Choose Optimal $\eta$ for Various Settings}
\label{alg:Opti_eta}
\SetKwData{Left}{left}\SetKwData{This}{this}\SetKwData{Up}{up}
\SetKwInOut{Input}{input}\SetKwInOut{Output}{output}
\Input{$\bm{Y}$, $\bm{S}^d$, $\bm{S}^t$, $\bm{U}$, $\bm{V}$, $k$, ${\cal C}$}
\Output{$\eta_2$, $\eta_3$, $\eta_4$}
$a_2, a_3, a_4 \leftarrow 0,0,0$ \;
$\bm{U}', \bm{V}' \leftarrow \bm{0}, \bm{0}$ \tcc*[r]{pseudo latent features}
\For{$\eta \in {\cal C}$}{
    \For{$x \leftarrow 1$ \KwTo $n_d$}{
        Compute $\bm{U}'_x$ according to Eq.\eqref{eq:U_x} with $\eta$  \;
    }   
    \For{$z \leftarrow 1$ \KwTo $n_t$}{
        Compute $\bm{V}'_z$ according to Eq.\eqref{eq:V_z} with $\eta$  \;
    }  
    $\hat{\bm{Y}}^2, \hat{\bm{Y}}^3, \hat{\bm{Y}}^4 \leftarrow \sigma_2(\bm{U}'\bm{V}^\top), \sigma_2(\bm{U}\bm{V}'^\top), \sigma_2(\bm{U}'\bm{V}'^\top)$ \;
    \For{$i \leftarrow 2$ \KwTo $4$}{
        $a = AUPR(\hat{\bm{Y}}^i,\bm{Y})$ \;
        \If{$a>a_i$}{
            $a_i, \eta_i \leftarrow a, \eta$ \;
        }
    }
}
\end{algorithm}

\subsection{MFAUC}
Given $\bm{Y}$ and its predictions $\hat{\bm{Y}}$, AUC, which counts the proportion of tuples where the interacting pair has higher prediction than the non-interacting one, is defined as:
\begin{equation}
AUC(\hat{\bm{Y}},\bm{Y}) = \frac{1}{|P|}\sum_{(ij,hl)\in P}\llbracket \hat{Y}_{ij} > \hat{Y}_{hl} \rrbracket 
\label{eq:AUROC}
\end{equation}
where $P_0 =\{(i,j)|Y_{ij}=0\} $ is the set of all non-interacting pairs, and $P=\{((i,j),(h,l))|(i,j) \in P_1, (h,l) \in P_0\}$ is the Cartesian product of $P_1$ and $P_0$. 
For simplicity, we use $(ij,hl)$ to denote $((i,j),(h,l))$ in sequel.

The maximization of the $AUC$ metric is equivalent to the minimization of the
following loss:
\begin{equation}
{\cal L}'_{AUC} = \sum_{(ij,hl)\in P}\llbracket \hat{Y}_{ij} \leq \hat{Y}_{hl} \rrbracket
\label{eq:loss_auc'}
\end{equation}
Although the discontinuous of indicator function would make the minimization of ${\cal L}_{AUC}$ render to the NP hard problem, we can derive a convex AUC loss which is consistent with Eq. \eqref{eq:loss_auc'} by replacing the indicator function with a convex surrogate function $\phi(\cdot)$~\cite{Gao2015OnOptimization,Gultekin2020MBA:Optimization}:  
\begin{equation}
{\cal L}_{AUC} = \sum_{(ij,hl)\in P} \phi( \zeta_{ijhl})
\end{equation}
where $\zeta_{ijhl}=\hat{Y}_{ij}-\hat{Y}_{hl}$.
In MFAUC, $\phi$ is defined as the logistic loss function, i.e. $\phi(x)=\log(1+\exp(-x))$, and its derivative is $\phi'(x)=-(1+\exp(x))^{-1}$. 
Besides, the prediction of MFAUC is defined as $\hat{\bm{Y}} = \bm{U}\bm{V}^\top $.
By replacing ${\cal L}(\hat{\bm{Y}},\bm{Y})$ with ${\cal L}_{AUC}$ defined in Eq.\eqref{eq:MF_obj}, we obtain the optimization problem of MFAUC:
\begin{equation}
\min_{\bm{U},\bm{V}} {\cal J}_{AUC}={\cal L}_{AUC} + {\cal R}(\bm{U},\bm{V})
\label{eq:MFAUC_obj}
\end{equation}

Similar to MFAUPR, we solve the Eq. \eqref{eq:MFAUC_obj} by updating $\bm{U}$ and $\bm{V}$ alternatively. The gradient of ${\cal L}_{AUC}$ w.r.t $\bm{U}_i$ is
\begin{equation}
\resizebox{.91\linewidth}{!}{$
\displaystyle
\nabla_{\bm{U}_i} {\cal L}_{AUC} =  \sum_{(ij,hl) \in P}  \phi'( \zeta_{ijhl} ) \bm{V}_{j}  -\sum_{(hl,ij)\in P}  \phi'(\zeta_{hlij}) \bm{V}_{j} 
$}
\label{eq:Lauc_Ui} 
\end{equation}
Because $|P|$ is quadratic to $n_dn_t$, the computational complexity of calculating $\nabla_{\bm{U}_i} {\cal L}_{AUC}$ is $O(rn_dn^2_t)$ which is costly even for hundreds of drugs and targets. To speed up the computation, we approximate $\nabla_{\bm{U}_i} {\cal L}_{AUC}$ by using a $n_dn_t$ sized set $P'$ sampled from $P$ to reduce the complexity to $O(rn_t)$ :
\begin{equation}
\nabla_{\bm{U}_i} {\cal L}_{AUC} \approx  \sum_{(ij,hl) \in P'}  \phi'( \zeta_{ijhl} ) \bm{V}_{j}  -\sum_{(hl,ij)\in P'}  \phi'(\zeta_{hlij}) \bm{V}_{j} 
\label{eq:Lauc_Ui2} 
\end{equation}
Based on Eq.\eqref{eq:Lauc_Ui2}, the gradient of ${\cal L}_{AUC}$ w.r.t $\bm{U}$ is defined as:
\begin{equation}
\nabla_{\bm{U}} {\cal L}_{AUC} = \left[\left(\nabla_{\bm{U}_1} {\cal L}_{AUC}\right)^\top,  \dotsc, \left(\nabla_{\bm{U}_{n_d}} {\cal L}_{AUC}\right)^\top  \right]^\top    
\end{equation}
Similarly, the gradient of ${\cal L}_{AUC}$ w.r.t $\bm{V}$ is
\begin{equation}
\nabla_{\bm{V}} {\cal L}_{AUC} = \left[\left(\nabla_{\bm{V}_1} {\cal L}_{AUC}\right)^\top,  \dotsc, \left(\nabla_{\bm{V}_{n_t}} {\cal L}_{AUC}\right)^\top  \right]^\top 
\end{equation}
\begin{equation}
\nabla_{\bm{V}_{j}} {\cal L}_{AUC} \approx \sum_{(ij,hl) \in P'}  \phi'( \zeta_{ijhl} )\bm{U}_i -\sum_{(hl,ij)\in P'}  \phi'(\zeta_{hlij})\bm{U}_i
\end{equation}
MFAUC employs the AdaGrad algorithm~\cite{Duchi2011AdaptiveOptimization} that adaptively chooses the update step based on previous gradients to diminish the influence of sampling on computing gradients of ${\cal L}_{AUC}$. 
The training of MFAUC is shown in Algorithm \ref{alg:MFAUC_train}, where $\circ a$ denotes the element wise exponentiation with the exponent being $a$. 
Besides, in the $\eta$ selection procedure, predictions of MFAUC are defined as $\hat{\bm{Y}}^2=\bm{U}'\bm{V}^\top, \hat{\bm{Y}}^3=\bm{U}\bm{V}'^\top, \hat{\bm{Y}}^4=\bm{U}'\bm{V}'^\top$ (line 8, Algorithm \ref{alg:Opti_eta}) and the $\eta$ with best AUC is the optimal value.  

\begin{algorithm}[h]
\caption{Training of MFAUC}
\label{alg:MFAUC_train}
\SetKwData{Left}{left}\SetKwData{This}{this}\SetKwData{Up}{up}
\SetKwInOut{Input}{input}\SetKwInOut{Output}{output}
\Input{$\bm{Y}$, $D$, $T$, $\bm{S}^d$ , $\bm{S}^t$, $\theta$, $k$, ${\cal C}$}
\Output{$\bm{U}$, $\bm{V}$, $\eta_2$, $\eta_3$, $\eta_4$}
Compute $\bm{L}^d$ and $\bm{L}^t$ based on sparsified $\bm{S}^d$ and $\bm{S}^t$ \;
Initialize $\bm{U}$ and $\bm{V}$ randomly\;
$\bm{G},\bm{H} \leftarrow \bm{0}, \bm{0}$ \tcc*[r]{accumulated gradients}
\Repeat{convergence}{
$\nabla_{\bm{U}} {\cal J}_{AUC} \leftarrow \nabla_{\bm{U}} {\cal L}_{AUC} + \nabla_{\bm{U}} {\cal R}(\bm{U},\bm{V})$ \;
$\bm{G} \leftarrow \bm{G} + \left(\nabla_{\bm{U}} {\cal J}_{AUC} \right)^{\circ2}$ \; 
$\bm{U} \leftarrow \bm{U} - \theta \nabla_{\bm{U}} {\cal J}_{AUC} \odot \bm{G}^{\circ-\frac{1}{2}}$ \; 
$\nabla_{\bm{V}}  {\cal J}_{AUC} \leftarrow \nabla_{\bm{V}} {\cal L}_{AUC} + \nabla_{\bm{V}} {\cal R}(\bm{U},\bm{V})$ \;
$\bm{H} \leftarrow \bm{H} + \left(\nabla_{\bm{V}} {\cal J}_{AUC} \right)^{\circ2}$ \; 
$\bm{V} \leftarrow \bm{V} - \theta \nabla_{\bm{V}} {\cal J}_{AUC} \odot \bm{H}^{\circ-\frac{1}{2}}$ \; 
}
Choose $\eta_2$, $\eta_3$, $\eta_4$ from ${\cal C}$ using Algorithm \ref{alg:Opti_eta}\; 
\end{algorithm}


\subsection{MF2A}
Both AUPR and AUC play a vital role in DTI prediction, while MFAUPR and MFAUC only optimize one measure and ignore the other. To fully exploit both area under the curve metrics, MF2A, as an ensemble of MFAUPR and MFAUC, integrates the predicted scores of the two MF models to obtain the final estimate for test drug-target pairs. Let $\hat{Y}^{AP}_{xz}$ and $\hat{Y}^{AC}_{xz}$ be the predicted scores of a test pair $(d_x, t_z)$ obtained by MFAUPR and MFAUC, respectively. The final prediction of $(d_x, t_z)$ is calculated as:
\begin{equation}
\hat{Y}_{xz} = \beta \hat{Y}^{AP}_{xz}+(1-\beta)\sigma_2(\hat{Y}^{AC}_{xz})
\end{equation}
where $\beta \in [0,1]$ adjusts the importance of MFAUPR and MFAUC to the final output, and the logistic function $\sigma_2$ converts $\hat{Y}^{AC}_{xz}$ into the same scale of $\hat{Y}^{AP}_{xz}$, i.e. (0,1).
The MFAUPR and MFAUC could be considered as two degenerate variants of MF2A with $\beta=0$ and $\beta=1$, respectively.



\subsection{Complexity Analysis}

For LIC, the complexity of calculating the local interaction consistency of one drug similarity matrix is $O(n_d+k|P_1|)$, and the overall computational cost of the drug similarity integration is $O(m_d(n_d^2+k|P_1|))$. Likewise, the complexity of combining target similarities is $O(m_t(n_t^2+k|P_1|))$.

The computational cost of MFAUPR mainly depends on the calculation of the gradients of the objective function in each iteration of the optimization. The complexities of computing gradients of ${\cal L}_{AP}$, $\nabla_{\bm{U}} {\cal R}(\bm{U},\bm{V})$ and $\nabla_{\bm{V}} {\cal R}(\bm{U},\bm{V})$ are $O(rn_bn_dn_t)$, $O(rn_d^2)$ and $O(rn_t^2)$, respectively. Considering that $n_bn_dn_t$ is usually larger than $n_d^2$ and $n_t^2$, the overall complexity of MFAUPR is $O(rn_bn_dn_t)$.
Regarding MFAUC, the complexity of calculating gradients of ${\cal L}_{AUC}$ using the sampled set $P'$ is $O(rn_dn_t)$, while the other steps have the same complexity with MFAUPR. Thus, the complexity of MFAUC is $O(r*\max\{n_d^2,n_t^2\})$.
For MF2A, as an ensemble of MFAUPR and MFAUC, its complexity is $O(r(n_bn_dn_t+\max\{n_d^2,n_t^2\}))$.



\section{Experiments}
In this section, the dataset and evaluation protocol used in the empirical study are described firstly. Then the experimental results are presented to show the effectiveness of proposed approaches. In addition, we investigate the convergence and parameter sensitivity of proposed MF models. Finally, the newly discovered interactions found by MF2A are reported.

\subsection{Dataset}
Two kinds of benchmark DTI datasets are used in this study. The first one is a collection of four golden standard datasets constructed by Yamanishi in 2007, of which each corresponds to a target protein family, namely Nuclear Receptors (NR), Ion Channel (IC), G-protein coupled receptors (GPCR), Enzyme (E)~\cite{Yamanishi2008PredictionSpaces}.  
Because the interactions in these datasets were found 14 years ago, we update them by adding newly discovered interactions between drugs and targets in these datasets recorded in the last version of KEGG~\cite{Kanehisa2017KEGG:Drugs}, DrugBank~\cite{Wishart2018DrugBank2018} and ChEMBL~\cite{Mendez2019ChEMBL:Data} databases. The second one is Luo's dataset, including 1923 DTIs between 708 drugs and 1512 protein targets retrieved from DrugBank 3.0~\cite{Luo2017AInformation}, which is usually used to evaluate the performance of network based methods due to its larger scale~\cite{Wan2019NeoDTI:Interactions,Xuan2021IntegratingPrediction}.
The characteristics of the updated golden standard and Luo's datasets are shown in Table \ref{tab:Dataset}, where the Sparsity denotes the proportion of non-zero values (interacting pairs) in the interaction matrix.


\begin{table}[h]
\centering
\caption{Characteristic of datasets}
\label{tab:Dataset}
\begin{tabular}{@{}cccccc@{}}
\toprule
Dataset & \#Drugs & \#Targets & \#Interactions & Sparsity\\ \midrule
NR & 54 & 26 & 166 & 0.118\\
GPCR & 223 & 95 & 1096 & 0.052\\
IC & 210 & 204 & 2331 & 0.054\\
E & 445 & 664 & 4256 & 0.014 \\ \midrule
Luo & 708 & 1512 & 1923 & 0.002 \\ \bottomrule
\end{tabular}
\end{table}

Multiple types of similarities are used to describe drug and target properties in various aspects. For updated golden standard datasets, we select four drug similarities and four target similarities with the higher local interaction consistency from~\cite{Nascimento2016APrediction}. For drugs, one similarity is derived from chemical structure using SIMCOMP~\cite{Hattori2003DevelopmentPathways} algorithm, while the other three are computed based on AERS-freq,  AERS-bit~\cite{Takarabe2012DrugApproach} and SIDER~\cite{Kuhn2016TheEffects} drug side effects profiles. For targets, three similarities based on amino acid sequences are calculated by using Normalized Smith–Waterman (SW) score~\cite{Smith1981IdentificationSubsequences} and spectrum kernel with 3-mers length (SPEC-k3) and 4-mers length (SPEC-k4)~\cite{Leslie2002TheClassification}, and the last one is the semantic similarity of gene ontology (GO) term annotation~\cite{Barrell2009TheResource}. Supplementary Table S1 and Section S1 summarize the information and local interaction consistency of all drug and target similarities in~\cite{Nascimento2016APrediction}.

The similarities for Luo's dataset are obtained from~\cite{Luo2017AInformation}. Specifically, we employ four types of drug similarities based on chemical structure, drug-drug interactions, drug side effects and drug-disease associations, as well as three kinds of target similarities based on target protein sequence, target-target interactions and target-disease associations in our study. The drug-drug interactions, drug side effects, protein–protein interactions, and disease related associations are extracted from DrugBank~\cite{Wishart2018DrugBank2018}, SIDER~\cite{Kuhn2016TheEffects}, HPRD~\cite{KeshavaPrasad2009HumanUpdate}, and Comparative Toxicogenomics Database~\cite{Davis2013The2013}, respectively.

\subsection{Experimental Setup}

Following~\cite{Pahikkala2015TowardPredictions}, four types of cross validation (CV) are conducted to examine the methods in the corresponding four prediction settings. In S1, the 10-fold CV on pairs is used, where one fold of pairs are removed for testing. In S2 (S3), the 10-fold CV on drugs (targets) is applied, where one drug (target) fold along with their corresponding rows (columns) in $\bm{Y}$ are separated for testing. The 3$\times$3-fold block-wise CV, which splits a drug fold and target fold along with the interactions between them for testing and uses interactions between remaining drugs and targets for training, is applied to S4. 
Furthermore, AUPR and AUC defined in Eq. \eqref{eq:AUPR} and \eqref{eq:AUROC} are used as evaluation measures. 

The proposed three methods are compared with fifteen DTI prediction approaches, including:
\begin{itemize}
\item A neighbourhood based method: WkNNIR~\cite{Liu2021Drug-TargetRecovery}
\item A bi-clustering tree model: EBiCTR~\cite{Pliakos2020Drug-targetReconstruction}
\item Five MF methods optimizing various losses: NRLMF~\cite{Liu2016NeighborhoodPrediction}, GRMF~\cite{Ezzat2017Drug-targetFactorization}, BRDTI~\cite{Peska2017Drug-targetApproach}, MSCMF~\cite{Zheng2013CollaborativeInteractions} and MLRE~\cite{Li2019DrugEmbedding} 
\item Three kernel methods: KSVM~\cite{Airola2018FastTrick}, KRLSM~\cite{Nascimento2016APrediction} and DRLSM~\cite{Ding2020IdentificationFusion}
\item Five network based methods: DDR~\cite{Olayan2018DDR:Approaches}, DTINet~\cite{Luo2017AInformation}, NeoDTI~\cite{Wan2019NeoDTI:Interactions}, GANDTI~\cite{Sun2020GraphInteractions}, DTIP~\cite{Xuan2021IntegratingPrediction}
\end{itemize}
The MF and kernel methods, along with DDR, are able to deal with all prediction settings. WkNNIR and EBiCTR are specifically designed to predict interactions involving new arriving drugs or/and targets (S2, S3, S4) and thus do not provide a direct way to handle S1. All network based methods, except for DDR, usually focus on predicting unseen interactions between known drugs and targets (S1). 
For comparing methods that can only handle a single type of drug and target similarity, namely WkNNIR, EBiCTR, NRLMF, GRMF, BRDTI and KSVM, the average of the corresponding multiple similarities are used as input. 

The parameters of the comparing methods are set or chosen based on the suggestions in respective articles. The number of neighbours $k$ is set as 5 in LIC, MFAUPR and MFAUC. For MFAUPR and MFAUC, learning rate $\theta$=0.1, candidate decay coefficient set ${\cal C}$=\{0.1,0.2,$\dotsc$,1.0\}, the number of iterations is 100, the dimension of embedding features $r$ is chosen from \{50,100\}, $\lambda_d$, $\lambda_t$ and $\lambda_r$ are chosen from \{$2^{-4}$,$2^{-2}$,$2^{0}$,$2^{2}$\}. The number of bins $n_b$ in MFAUPR is chosen from \{11,16,21,26,31\}, and the trade-off ensemble weight $\beta$ in MF2A is chosen from $\{0,0.01,\cdots,1.0\}$.   
The codes and datasets used in this study are available at \url{https://github.com/intelligence-csd-auth-gr/DTI_MF2A}.

\subsection{Experimental Results}


We firstly compare the proposed MFAUPR, MFAUC and MF2A with eleven competitors applicable to more than one prediction settings, namely neighbourhood, tree, MF and kernel based approaches as well as DDR, using the updated golden standard datasets.
The results in terms of AUPR and AUC are shown in Tables \ref{tab:AUPR_Results} and \ref{tab:AUC_Results} respectively, where the second last row lists the average ranks across all settings, and the marks $\circ$, $\bullet$ and $\star$ in the last row indicate whether MFAUPR, MFAUC and MF2A are significantly better than the corresponding method respectively, according to the Wilcoxon signed-rank test with Bergman-Hommel’s correction \cite{Benavoli2016} at 5\% level on the results of all prediction settings. 

At first, we focus on the proposed three MF methods. 
MFAUPR and MFAUC are the second-best methods, only outperformed by MF2A, and achieve significant wins to the other twelve comparing approaches in terms of the corresponding metric they optimize, respectively. This verifies that optimizing surrogate AUPR and AUC losses is more effective than leveraging other losses to distinguish interacting pairs from numerous non-interacting ones.
However, they do not perform outstandingly in terms of the other metric (AUC for MFAUPR and AUPR for MFAUC), being the fifth method and even inferior to two competitors. This is an expected outcome, as MFAUPR and MFAUC optimize one metric and ignore the other, and optimizing AUPR is not always guaranteed to optimize AUC and vice versa~\cite{Davis2006TheCurves}.
MF2A is the best method in all prediction settings and significantly outperforms all other competitors, including MFAUPR and MFAUC, in terms of both evaluation metrics. The success of MF2A is mainly due to the ensembling of MFAUPR and MFAUC, which not only simultaneously optimizes both metrics but also combines two diverse models to reduce the prediction error. 

Concerning the competitors, WkNNIR and GRMF are the third and fourth methods, following MF2A and MFAUPR, in terms of AUPR, and perform moderately in terms of AUC. This is because they exploit the more informative recovery interactions inferred from neighbourhoods, which essentially improves the importance of interacting pairs in the prediction model. 
On the other hand, EBiCTR that incorporates the matrix factorization based interaction reconstruction is worse than WkNNIR and GRMF. One possible reason for this outcome is that matrix factorization replenishes interactions in the global view, losing the local properties provided by the neighbourhood. 
NRLMF is the third best method in terms of AUC and achieves mediocre results in terms of AUPR, since it optimizes the logistic loss and imposes a higher weight on interacting pairs.
The approaches that directly work on square loss without emphasizing interaction pairs, including DRLSM, KRLSM, MSCMF, MLRE, and DDR, are among the worst methods due to their bias to non-interacting pairs. 
KSVM and BRDTI are better than square loss-based methods in most cases.
KSVM optimizes the square hinge loss, which relies on fewer boundary pairs and therefore is less sensitive to the global imbalanced class distribution than the square loss, especially when the amounts of interacting and non-interacting pairs near the hyperplane are similar.
In addition, BRDTI achieves the third-best AUC result in S2, but performs worse in other settings.
The drug-based ranking loss optimized by BRDTI formulates the ideal order of targets for each drug, which is suitable for S2 that aims to predict interactions between a fixed target set and new drugs. 
Nevertheless, random pairs are masked as the test set in S1, and interactions for new targets should be predicted in S3 and S4, leading to the incomplete per-drug target sequence used to train the model. Therefore, BRDTI collapses in those three settings.


\begin{table*}[h]
\centering
\setlength{\tabcolsep}{2pt}
\caption{AUPR results on updated golden standard datasets, where the top three results are hilighted by \textbf{bold}, {\color[HTML]{FF0000}red} and {\color[HTML]{4472C4} blue} font respectively}
\label{tab:AUPR_Results}
\begin{tabular}{ccccccccccccc|ccc}
\toprule
Setting & Dataset & WkNNIR & EBiCTR & NRLMF & GRMF & BRDTI & MSCMF & MLRE & KSVM & KRLSM & DRLSM & DDR & MFAUPR & MFAUC & MF2A \\ \hline
 & NR & - & - & 0.64 & 0.646 & 0.573 & 0.628 & 0.534 & {\color[HTML]{FF0000} 0.664} & 0.642 & 0.64 & 0.484 & {\color[HTML]{FF0000} 0.664} & 0.644 & \textbf{0.673} \\
 & GPCR & - & - & 0.86 & 0.85 & 0.746 & 0.844 & 0.34 & 0.845 & 0.835 & 0.841 & 0.738 & {\color[HTML]{4472C4} 0.869} & \textbf{0.873} & {\color[HTML]{FF0000} 0.87} \\
 & IC & - & - & 0.934 & 0.919 & 0.75 & {\color[HTML]{4472C4} 0.936} & 0.503 & 0.914 & 0.914 & 0.931 & 0.898 & {\color[HTML]{FF0000} 0.94} & 0.931 & \textbf{0.943} \\
\multirow{-4}{*}{S1} & E & - & - & {\color[HTML]{4472C4} 0.843} & 0.826 & 0.714 & 0.818 & 0.282 & 0.78 & 0.811 & 0.836 & 0.698 & 0.835 & {\color[HTML]{FF0000} 0.846} & \textbf{0.858} \\ \cline{3-16} 
\multicolumn{2}{c}{$AveRank$} & - & - & 4.63 & 5.5 & 10.25 & 6.5 & 11.75 & 6.5 & 7.88 & 6.25 & 11 & {\color[HTML]{FF0000} 3.13} & {\color[HTML]{4472C4} 3.38} & \textbf{1.25} \\ \hline
 & NR & 0.562 & 0.545 & 0.547 & 0.564 & 0.545 & 0.531 & 0.403 & 0.543 & {\color[HTML]{4472C4} 0.57} & 0.477 & 0.317 & \textbf{0.58} & 0.541 & {\color[HTML]{FF0000} 0.578} \\
 & GPCR & {\color[HTML]{4472C4} 0.54} & 0.515 & 0.508 & 0.522 & 0.534 & 0.472 & 0.218 & 0.518 & 0.532 & 0.373 & 0.184 & {\color[HTML]{FF0000} 0.545} & 0.534 & \textbf{0.551} \\
 & IC & {\color[HTML]{4472C4} 0.491} & 0.49 & 0.479 & 0.49 & 0.476 & 0.379 & 0.22 & 0.474 & 0.466 & 0.318 & 0.133 & {\color[HTML]{FF0000} 0.492} & 0.474 & \textbf{0.495} \\
\multirow{-4}{*}{S2} & E & {\color[HTML]{4472C4} 0.405} & 0.368 & 0.389 & 0.381 & 0.358 & 0.288 & 0.095 & 0.37 & 0.376 & 0.267 & 0.096 & {\color[HTML]{FF0000} 0.415} & 0.4 & \textbf{0.422} \\ \cline{3-16} 
\multicolumn{2}{c}{$AveRank$} & {\color[HTML]{4472C4} 3.5} & 7.5 & 6.75 & 5.38 & 7.25 & 11 & 13.25 & 8.38 & 6.5 & 12 & 13.75 & {\color[HTML]{FF0000} 1.75} & 6.75 & \textbf{1.25} \\ \hline
 & NR & {\color[HTML]{4472C4} 0.56} & 0.413 & \textbf{0.519} & 0.534 & 0.468 & 0.505 & 0.397 & 0.532 & 0.546 & 0.498 & 0.275 & {\color[HTML]{FF0000} 0.587} & 0.553 & \textbf{0.588} \\
 & GPCR & 0.774 & 0.733 & 0.729 & {\color[HTML]{4472C4} 0.777} & 0.634 & 0.69 & 0.257 & 0.753 & 0.757 & 0.67 & 0.212 & {\color[HTML]{FF0000} 0.786} & 0.772 & \textbf{0.787} \\
 & IC & 0.861 & 0.835 & 0.838 & \textbf{0.865} & 0.758 & 0.827 & 0.43 & \textbf{0.865} & 0.855 & 0.795 & 0.472 & 0.86 & 0.858 & {\color[HTML]{4472C4} 0.863} \\
\multirow{-4}{*}{S3} & E & {\color[HTML]{FF0000} 0.728} & {\color[HTML]{4472C4} 0.723} & 0.711 & 0.721 & 0.682 & 0.623 & 0.289 & 0.628 & 0.712 & 0.576 & 0.16 & 0.719 & 0.718 & \textbf{0.731} \\ \cline{3-16} 
\multicolumn{2}{c}{$AveRank$} & {\color[HTML]{4472C4} 3.25} & 8 & 8.25 & 3.63 & 11 & 10 & 13.25 & 6.38 & 6.25 & 11 & 13.75 & {\color[HTML]{FF0000} 3.5} & 5.25 & \textbf{1.5} \\ \hline
 & NR & {\color[HTML]{FF0000} 0.309} & 0.214 & 0.278 & \textbf{0.313} & {\color[HTML]{4472C4} 0.292} & 0.273 & 0.226 & 0.281 & 0.236 & 0.223 & 0.116 & 0.284 & 0.285 & 0.289 \\
 & GPCR & 0.393 & 0.36 & 0.331 & 0.385 & 0.318 & 0.323 & 0.183 & 0.39 & 0.077 & 0.264 & 0.048 & {\color[HTML]{FF0000} 0.404} & {\color[HTML]{4472C4} 0.401} & \textbf{0.407} \\
 & IC & 0.339 & 0.337 & 0.327 & 0.338 & 0.324 & 0.194 & 0.172 & 0.337 & 0.086 & 0.246 & 0.053 & {\color[HTML]{FF0000} 0.351} & {\color[HTML]{4472C4} 0.344} & \textbf{0.352} \\
\multirow{-4}{*}{S4} & E & {\color[HTML]{4472C4} 0.228} & 0.188 & 0.221 & 0.22 & 0.208 & 0.074 & 0.077 & 0.099 & 0.063 & 0.086 & 0.014 & {\color[HTML]{FF0000} 0.234} & 0.215 & \textbf{0.235} \\ \cline{3-16} 
\multicolumn{2}{c}{$AveRank$} & {\color[HTML]{4472C4} 3.25} & 8.63 & 7 & 4.25 & 7.25 & 10.25 & 11.5 & 6.88 & 12.25 & 10.75 & 14 & {\color[HTML]{FF0000} 3} & 4.25 & \textbf{1.75} \\ \hline
\multicolumn{2}{c}{} & {\color[HTML]{4472C4} 3.33} & 8.04 & 6.66 & 4.69 & 8.94 & 9.44 & 12.44 & 7.04 & 8.22 & 10 & 13.13 & {\color[HTML]{FF0000} 2.85} & 4.91 & \textbf{1.44} \\ 
&  & $\circ$$\star$ & $\circ$$\bullet$$\star$ & $\circ$$\bullet$$\star$ & $\circ$$\star$ & $\circ$$\bullet$$\star$ & $\circ$$\bullet$$\star$ & $\circ$$\bullet$$\star$ & $\circ$$\bullet$$\star$ & $\circ$$\bullet$$\star$ & $\circ$$\bullet$$\star$ & $\circ$$\bullet$$\star$ & $\star$ & $\circ$$\star$ & \\ \bottomrule
\end{tabular}
\end{table*}

\begin{table*}[ht]
\centering
\setlength{\tabcolsep}{2pt}
\caption{AUC results on updated golden standard datasets, where the top three results are hilighted by \textbf{bold}, {\color[HTML]{FF0000}red} and {\color[HTML]{4472C4} blue} font respectively}
\label{tab:AUC_Results}
 \begin{tabular}{ccccccccccccc|ccc}
\toprule
Setting & Dataset & WkNNIR & EBiCTR & NRLMF & GRMF & BRDTI & MSCMF & MLRE & KSVM & KRLSM & DRLSM & DDR & MFAUPR & MFAUC & MF2A \\ \hline
 & NR & - & - & {\color[HTML]{4472C4} 0.882} & 0.874 & 0.849 & {\color[HTML]{4472C4} 0.882} & 0.835 & 0.877 & 0.879 & 0.879 & 0.826 & {\color[HTML]{4472C4} 0.882} & 0.879 & \textbf{0.884} \\
 & GPCR & - & - & 0.972 & 0.97 & 0.962 & 0.962 & 0.87 & 0.969 & 0.971 & 0.972 & 0.922 & {\color[HTML]{FF0000} 0.977} & {\color[HTML]{4472C4} 0.975} & \textbf{0.978} \\
 & IC & - & - & \textbf{0.989} & 0.983 & 0.946 & 0.982 & 0.926 & 0.98 & 0.981 & 0.986 & 0.971 & 0.986 & {\color[HTML]{4472C4} 0.988} & \textbf{0.989} \\
\multirow{-4}{*}{S1} & E & - & - & {\color[HTML]{4472C4} 0.981} & 0.971 & 0.97 & 0.961 & 0.922 & 0.959 & 0.964 & 0.973 & 0.921 & 0.97 & \textbf{0.983} & \textbf{0.983} \\ \cline{3-16} 
\multicolumn{2}{c}{$AveRank$} & - & - & {\color[HTML]{FF0000} 3} & 6.75 & 9.25 & 7.13 & 11.5 & 8.75 & 7 & 4.75 & 11.25 & 4 & {\color[HTML]{4472C4} 3.38} & \textbf{1.25} \\ \hline
 & NR & 0.825 & 0.811 & 0.826 & 0.83 & \textbf{0.833} & 0.802 & 0.752 & 0.821 & 0.831 & 0.789 & 0.666 & 0.827 & {\color[HTML]{4472C4} 0.832} & \textbf{0.833} \\
 & GPCR & 0.914 & {\color[HTML]{4472C4} 0.922} & 0.913 & 0.907 & 0.91 & 0.882 & 0.832 & 0.883 & 0.867 & 0.856 & 0.588 & 0.919 & {\color[HTML]{FF0000} 0.923} & \textbf{0.924} \\
 & IC & {\color[HTML]{4472C4} 0.826} & 0.816 & 0.825 & 0.811 & 0.82 & 0.783 & 0.758 & 0.819 & 0.796 & 0.773 & 0.541 & 0.797 & {\color[HTML]{FF0000} 0.827} & \textbf{0.828} \\
\multirow{-4}{*}{S2} & E & 0.877 & \textbf{0.906} & 0.858 & 0.847 & 0.869 & 0.835 & 0.795 & 0.842 & 0.799 & 0.81 & 0.535 & 0.847 & {\color[HTML]{4472C4} 0.891} & {\color[HTML]{FF0000} 0.892} \\ \cline{3-16} 
\multicolumn{2}{c}{$AveRank$} & 5 & 5.25 & 5.75 & 7.13 & {\color[HTML]{4472C4} 4.63} & 10.5 & 13 & 8.25 & 9.25 & 11.75 & 14 & 6.63 & {\color[HTML]{FF0000} 2.5} & \textbf{1.38} \\ \hline
 & NR & {\color[HTML]{4472C4} 0.82} & 0.766 & \textbf{0.825} & 0.819 & 0.783 & 0.786 & 0.733 & 0.785 & 0.798 & 0.759 & 0.645 & 0.801 & {\color[HTML]{FF0000} 0.824} & 0.819 \\
 & GPCR & 0.952 & 0.926 & 0.946 & 0.955 & 0.936 & 0.902 & 0.855 & 0.93 & 0.917 & 0.901 & 0.576 & {\color[HTML]{4472C4} 0.956} & \textbf{0.96} & \textbf{0.96} \\
 & IC & 0.958 & 0.952 & 0.96 & 0.958 & 0.952 & 0.941 & 0.872 & 0.949 & 0.942 & 0.938 & 0.758 & {\color[HTML]{4472C4} 0.964} & {\color[HTML]{FF0000} 0.965} & \textbf{0.967} \\
\multirow{-4}{*}{S3} & E & 0.935 & 0.931 & {\color[HTML]{4472C4} 0.936} & 0.92 & 0.922 & 0.881 & 0.895 & 0.875 & 0.883 & 0.913 & 0.511 & 0.932 & \textbf{0.944} & \textbf{0.944} \\ \cline{3-16} 
\multicolumn{2}{c}{$AveRank$} & 4.38 & 8.38 & {\color[HTML]{4472C4} 3.5} & 5.5 & 7.88 & 10.5 & 12.25 & 9.75 & 9.5 & 11.25 & 14 & 4.25 & \textbf{1.75} & {\color[HTML]{FF0000}2.13} \\ \hline
 & NR & 0.637 & 0.587 & {\color[HTML]{FF0000} 0.656} & 0.637 & \textbf{0.683} & 0.597 & 0.62 & 0.589 & 0.592 & 0.56 & 0.484 & 0.648 & 0.647 & {\color[HTML]{4472C4} 0.649} \\
 & GPCR & {\color[HTML]{4472C4} 0.871} & 0.855 & 0.866 & 0.853 & 0.855 & 0.798 & 0.792 & 0.859 & 0.405 & 0.786 & 0.345 & {\color[HTML]{4472C4} 0.871} & {\color[HTML]{FF0000} 0.885} & \textbf{0.886} \\
 & IC & 0.774 & 0.751 & {\color[HTML]{4472C4} 0.775} & 0.768 & 0.752 & 0.658 & 0.664 & 0.754 & 0.498 & 0.718 & 0.409 & 0.757 & \textbf{0.776} & \textbf{0.776} \\
\multirow{-4}{*}{S4} & E & {\color[HTML]{4472C4} 0.819} & 0.812 & 0.799 & 0.777 & 0.779 & 0.695 & 0.731 & 0.705 & 0.453 & 0.734 & 0.44 & 0.798 & \textbf{0.823} & {\color[HTML]{FF0000} 0.821} \\ \cline{3-16} 
\multicolumn{2}{c}{$AveRank$} & 4.25 & 8.13 & 3.75 & 7.13 & 5.88 & 10.75 & 10 & 8.75 & 12.25 & 11 & 14 & 4.88 & {\color[HTML]{FF0000} 2.38} & \textbf{1.88} \\ \hline
\multicolumn{2}{c}{} & 4.54 & 7.25 & {\color[HTML]{4472C4} 4} & 6.63 & 6.91 & 9.72 & 11.69 & 8.88 & 9.5 & 9.69 & 13.31 & 4.94 & {\color[HTML]{FF0000} 2.5} & \textbf{1.66} \\
&  & $\bullet$$\star$ & $\bullet$$\star$ & $\bullet$$\star$ & $\bullet$$\star$ & $\bullet$$\star$ & $\circ$$\bullet$$\star$ &  $\circ$$\bullet$$\star$ & $\circ$$\bullet$$\star$ & $\circ$$\bullet$$\star$ & $\circ$$\bullet$$\star$ & $\circ$$\bullet$$\star$ & $\bullet$$\star$ & $\star$ &
\\ \bottomrule
\end{tabular}
\end{table*}


\begin{table*}[h]
\centering
\caption{AUC and AUPR results on Luo's dataset in S1, where the top three results are hilighted by \textbf{bold}, {\color[HTML]{FF0000}red} and {\color[HTML]{4472C4} blue} font respectively}
\label{tab:result_luo}
\begin{tabular}{@{}ccccccccc@{}}
\toprule
Metric & DDR & DTINet & NeoDTI & GANDTI & DTIP & MFAUPR & MFAUC & MF2A \\ \midrule
AUPR & 0.05 (8) & 0.175 (6) & 0.539 (4) & 0.069 (7) & 0.399 (5) & {\color[HTML]{FF0000} 0.632 (2)} & {\color[HTML]{4472C4} 0.6 (3)} & \textbf{0.653 (1)} \\
AUC & 0.879 (7) & 0.921 (5) & 0.911 (6) & 0.867 (8) & \textbf{0.981 (1)} & 0.925 (4) & {\color[HTML]{4472C4} 0.965 (3)} & {\color[HTML]{FF0000} 0.966 (2)} \\ \bottomrule
\end{tabular} 
\end{table*}


Moreover, we compare the prediction performance of our approaches with five network based methods on Luo's dataset. Table \ref{tab:result_luo} lists the AUPR and AUC results, where only S1 is considered because most network based methods cannot handle other prediction settings.
As can be seen, MF2A, MFAUPR and MFAUC are the top three methods in terms of AUPR, which achieve 21.2\%, 17.3\% and 11.3\% improvements over the best competitor (NeoDTI), respectively.
With respect to AUC, MF2A achieves the second best result, which is 1.5\% lower than the best competitor DTIP. Nevertheless, DTIP performs worse in terms of AUPR, indicating that it heavily emphasizes the AUC performance at the cost of deteriorating the AUPR result. MFAUC and MFAUPR come next, outperforming the second best comparing method (DTINet) by 4.8\% and 0.4\%, respectively.


Overall, all the above experimental results confirm that MF2A achieves outstanding performance in terms of both AUPR and AUC. MFAUPR and MFAUC perform well in terms of the corresponding metric they optimize, but their performance declines in terms of the other metric.

\subsection{Effectiveness of LIC}
To validate the effectiveness of the proposed LIC, we compare it with four similarity integration approaches used in the DTI prediction task, such as AVE, KA~\cite{Nascimento2016APrediction}, HSIC~\cite{Ding2020IdentificationFusion}, and SNF~\cite{Olayan2018DDR:Approaches}. In this experiment, the LIC embedded in MF2A is replaced by other similarity integration competitors, and the results in S4 are shown in Table \ref{tab:Various_SimI}. 

As can be seen, LIC performs best in most cases, followed by KA and HSIC. 
This demonstrates that the local interaction consistency used in LIC distinguishes high-quality similarities better than the global one leveraged by KA and HSIC. 
In addition, HSIC is the best method on the NR dataset that contains only tens of drugs and targets. This is because the local and global interaction consistency provide similar information in this small-sized dataset.
AVE and SNF that totally ignore the interaction information are the two worst approaches. Furthermore, SNF is usually inferior to AVE. We speculate that the similarity with larger values, which dominates the similarity fusion procedure in SNF, may fail to supply enough beneficial information for DTI prediction.
Similar results are also observed in the other three prediction settings (See supplementary Table S3).



\begin{table}[]
\centering
\caption{Results of MF2A using difference similarity integration methods in S4}
\label{tab:Various_SimI}
\begin{tabular}{ccccccc}
\toprule
Metric & Dataset & Ave & KA & HSIC & SNF & LIC \\ \hline
\multirow{4}{*}{AUPR} & NR & 0.294 & 0.284 & \textbf{0.297} & 0.238 & 0.289 \\
 & GPCR & 0.4 & 0.401 & 0.401 & 0.379 & \textbf{0.407} \\
 & IC & 0.34 & 0.348 & 0.341 & 0.282 & \textbf{0.352} \\
 & E & 0.213 & 0.225 & 0.215 & 0.218 & \textbf{0.235} \\
\multicolumn{2}{c}{\textit{AveRank}} & 3.75 & 2.63 & 2.63 & 4.5 & \textbf{1.5} \\ \hline
\multirow{4}{*}{AUC} & NR & 0.658 & 0.649 & \textbf{0.66} & 0.622 & 0.649 \\
 & GPCR & 0.883 & 0.884 & 0.883 & 0.874 & \textbf{0.886} \\
 & IC & 0.772 & 0.773 & 0.771 & 0.758 & \textbf{0.776} \\
 & E & 0.813 & 0.817 & 0.816 & 0.802 & \textbf{0.821} \\
\multicolumn{2}{c}{\textit{AveRank}} & 3.13 & 2.38 & 2.88 & 5 & \textbf{1.63} \\ \bottomrule
\end{tabular}
\end{table}

\subsection{Convergence and Parameter Analysis}
In this section, we study the convergence and parameter sensitivity of proposed methods.

\begin{figure}[h]
\centering
\subfloat[MFAUPR]{\includegraphics[width=0.23\textwidth]{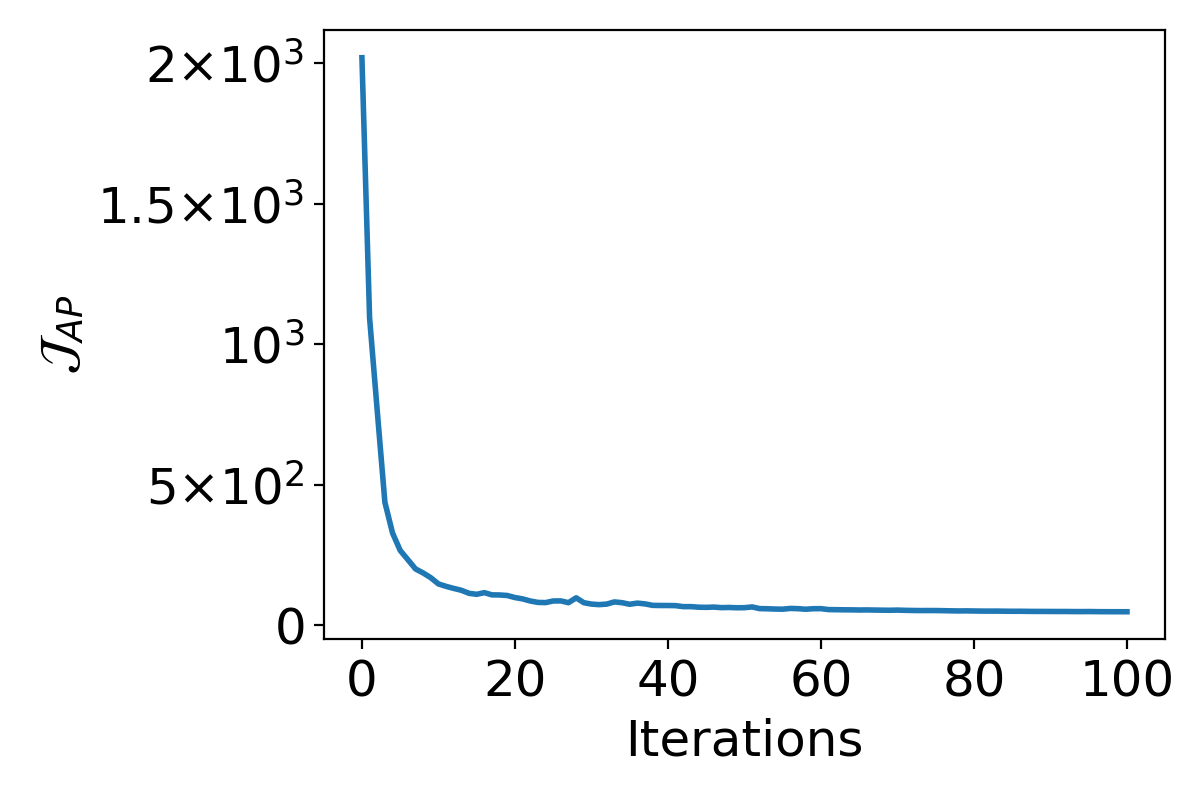}}
\hspace{1em}
\subfloat[MFAUC]{\includegraphics[width=0.23\textwidth]{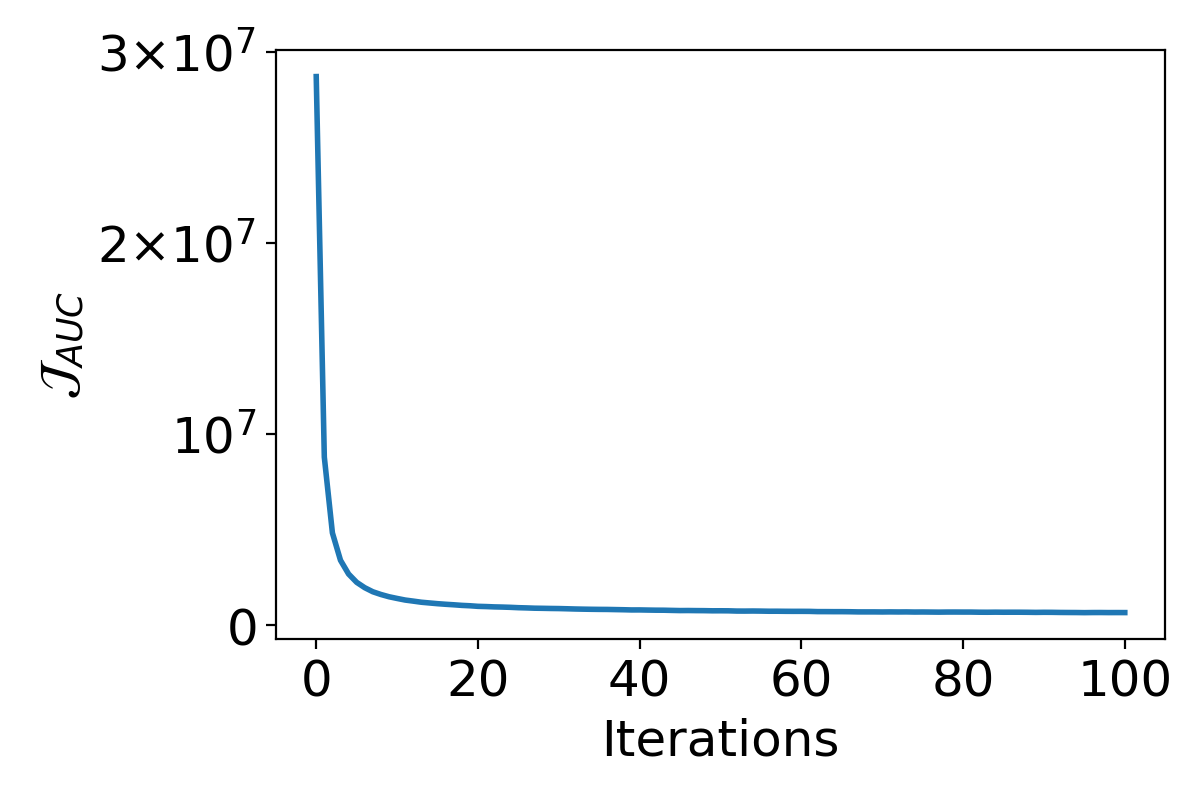}}
\caption{Convergence of MFAUPR and MFAUC on IC dataset in S1} 
\label{fig:Convergence}
\end{figure}

Figure \ref{fig:Convergence} shows the convergence curves of MFAUPR and MFAUC on IC dataset in S1. The values of the objective functions drop sharply in the first 10 iterations and then gradually decrease to a stationary state, which demonstrates the convergence of the proposed methods.

\begin{figure*}[h]
\centering
\subfloat[MFAUPR]{\includegraphics[width=0.25\textwidth]{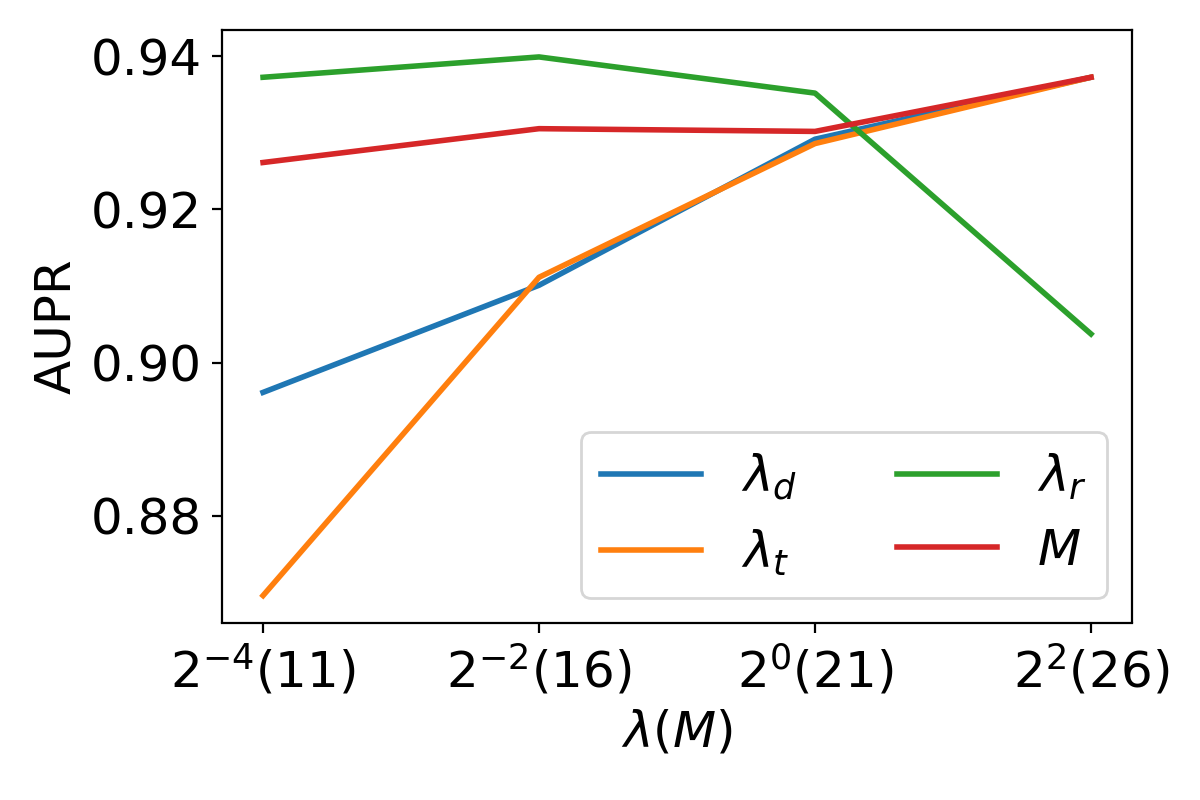}}
\hspace{1em}
\subfloat[MFAUC]{\includegraphics[width=0.25\textwidth]{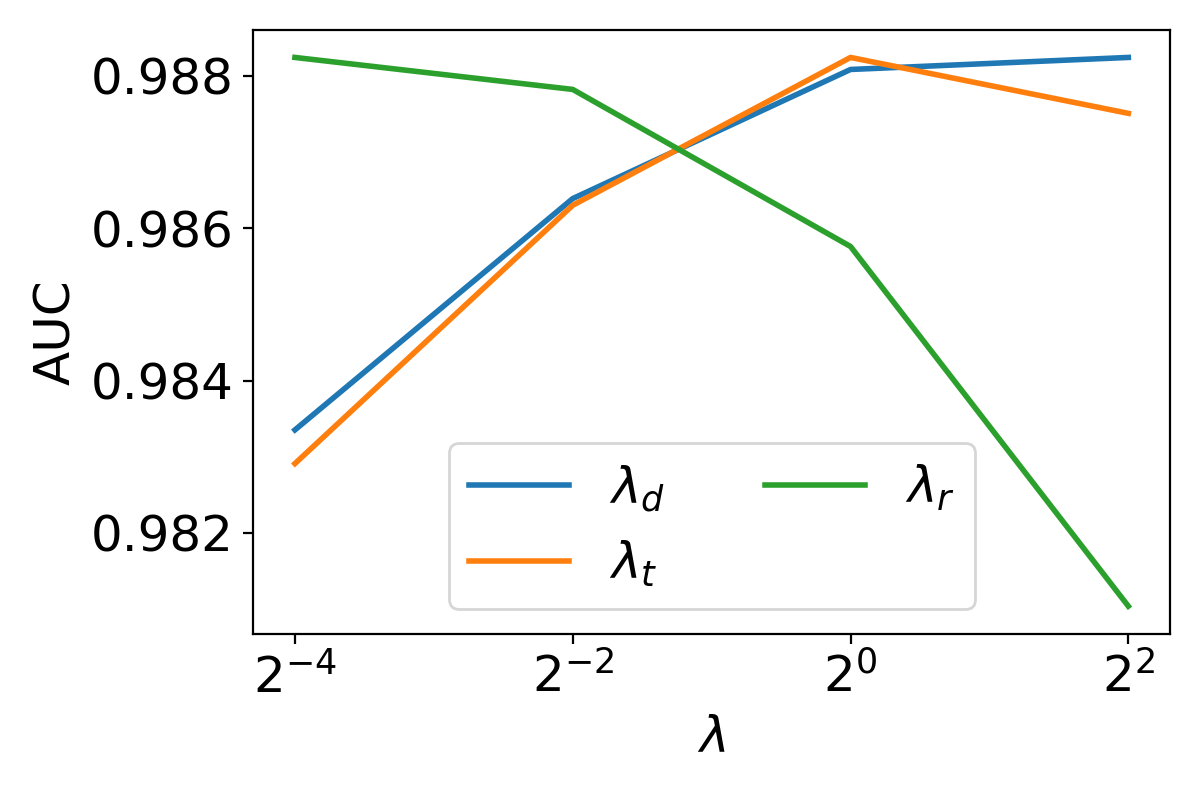}}
\hspace{1em}
\subfloat[MF2A]{\includegraphics[width=0.25\textwidth]{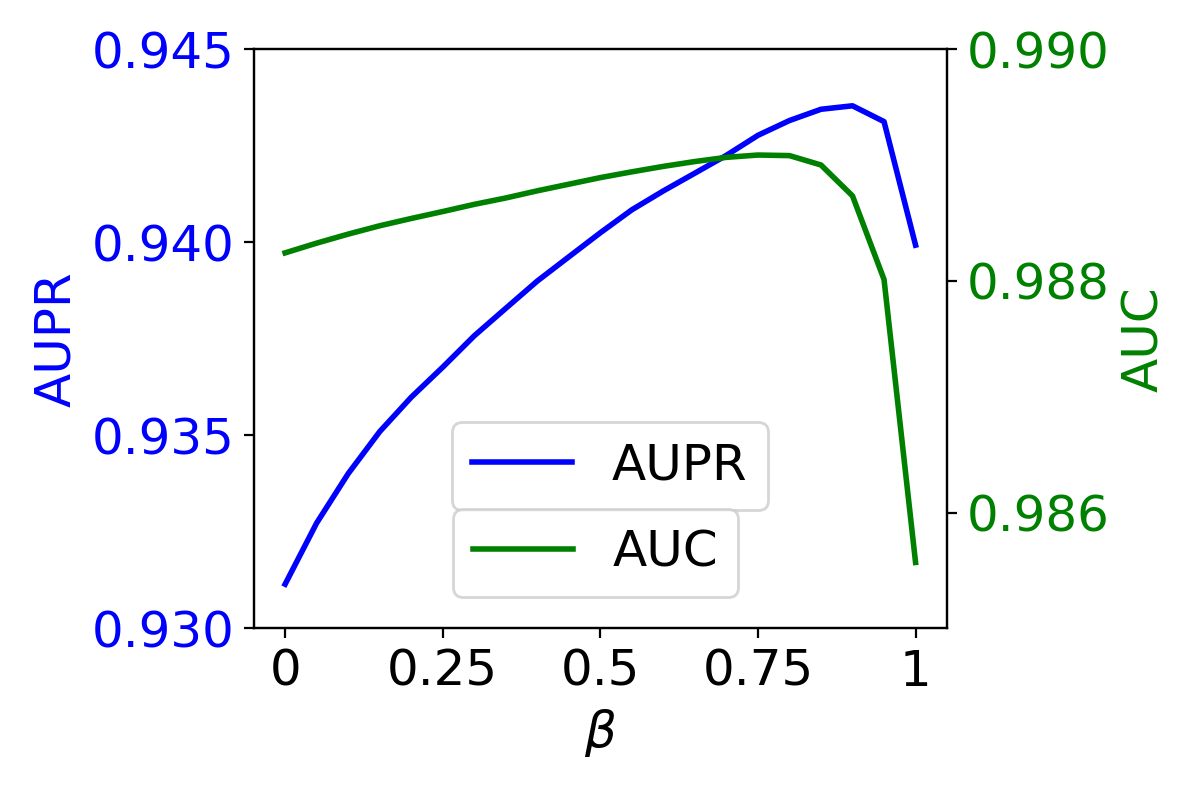}
\label{fig:VarParams_MF2A}}
\caption{Results of MFAUPR (a), MFAUC (b) and MF2A (c) with various parameter settings on IC dataset in S1} 
\label{fig:VarParams}
\end{figure*}

Figure \ref{fig:VarParams} presents the performance of proposed methods with various parameters settings on IC dataset in S1. 
With respect to the regularization coefficients in MFAUPR and MFAUC, the optimal values of $\lambda_d$ and $\lambda_t$ that relate to graph regularization are usually higher than the best option of $\lambda_r$ which controls Tikhonov regularization. This implies that graph regularization, which retains the similarity based topological structure in the latent feature space, is more important than the Tikhonov regularization that avoids the excessively large latent feature values. 
Regarding the number of bins ($n_b$) in MFAUPR, a larger value provides a more fine grain division for predictions, leading to better performance. 
The influence of ensemble weight $\beta$ in MF2A is shown in Figure \ref{fig:VarParams_MF2A}. When $\beta=0$, MF2A degenerates to MFAUC, so the AUPR performance is lower. 
With increasing $\beta$, both AUPR and AUC performances improve, as both MFAUPR and MFAUC are aggregated in MF2A. 
However, when $\beta$ is too large, the performances deteriorate, as MF2A relies too much on MFAUPR.

\begin{figure}[h]
\centering
\subfloat[AUPR]{\includegraphics[width=0.23\textwidth]{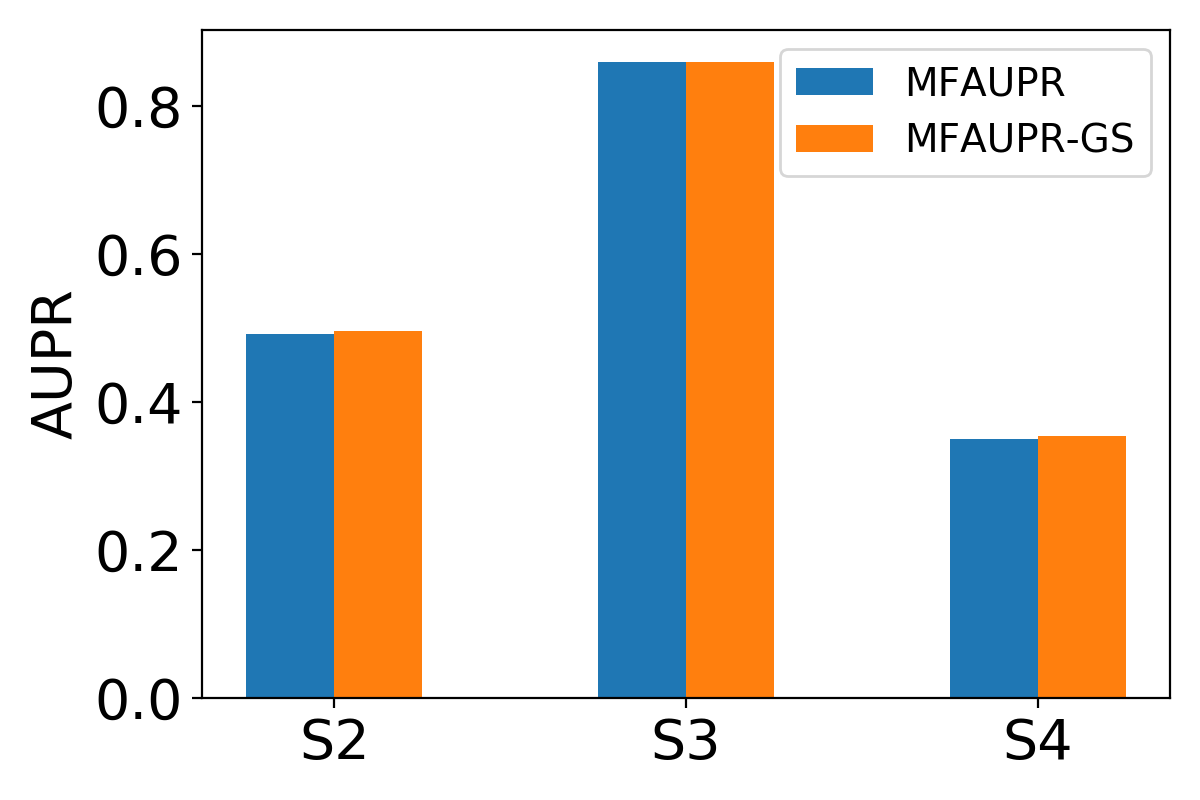}}
\hspace{1em}
\subfloat[Training time ratio (MFAUPR-GS : MFAUPR)]{\includegraphics[width=0.23\textwidth]{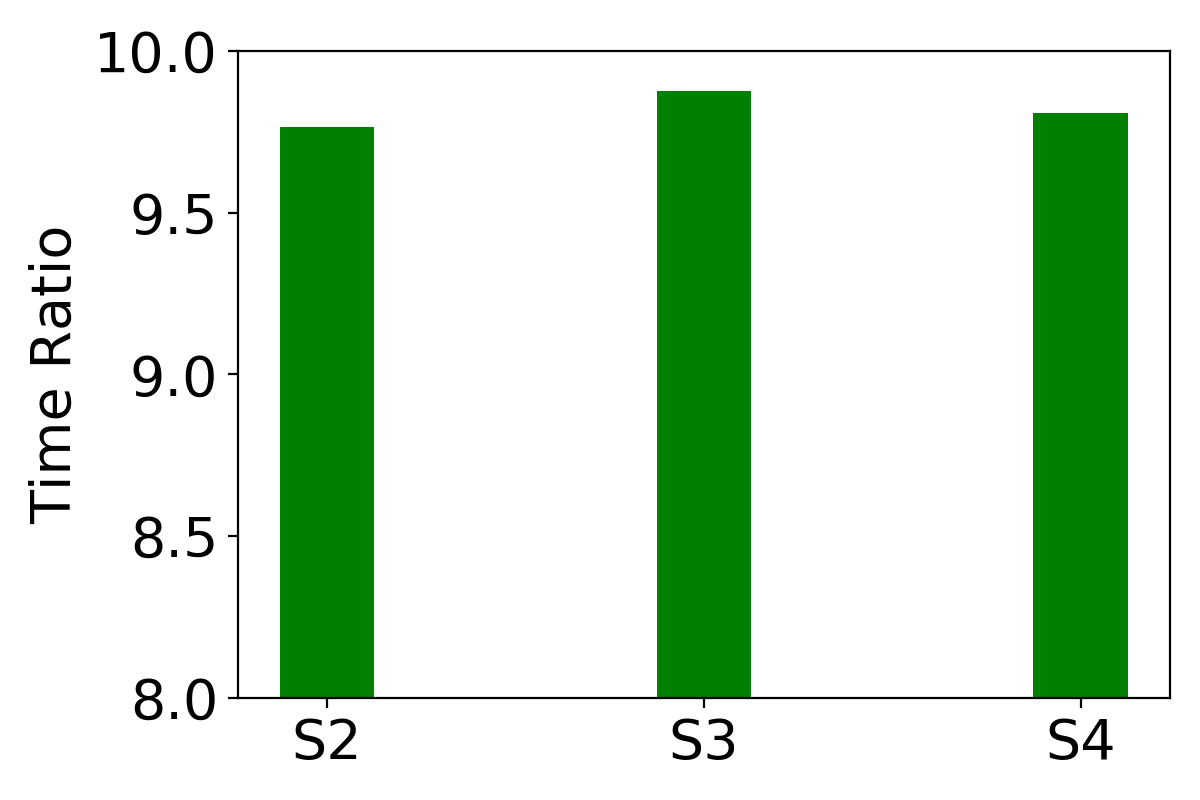}}
\caption{Comparison of MFAUPR and MFAUPR-GS on IC dataset} 
\label{fig:Optimal_eta_MFAUPR}
\end{figure}

The decay coefficient $\eta$ is a crucial parameter for the computation of test drugs (targets) features. To examine the effectiveness of the proposed optimal $\eta$ selection strategy, we compare the AUPR results and the total training time of MFAUPR with a variant of MFAUPR (denoted as MFAUPR-GS) that tunes $\eta$ via grid search instead of Algorithm \ref{alg:Opti_eta}. In MFAUPR-GS, 10 models are trained for 10 $\eta$ candidates respectively, and the $\eta$ value that achieves the best AUPR on the test set is chosen.  
As shown in Figure \ref{fig:Optimal_eta_MFAUPR}, MFAUPR achieves similar performance with MFAUPR-GS with more than 9.5 times speed-up in S2, S3 and S4 that involve predicting interactions for new drugs or targets. This demonstrates the effectiveness of our $\eta$ selection strategy to pick up the ideal parameter setting.
A similar phenomenon can be observed for MFAUC (See supplementary Figure S2).

\subsection{Discovery of Novel Interactions}
In this section, we examine the ability of the best proposed approach, MF2A, to discover recently confirmed DTIs that are not known in the datasets. For the four golden standard datasets, their \textit{original} versions~\cite{Yamanishi2008PredictionSpaces} are used in this part, because the updated ones have already contained all newly found DTIs collected from up-to-date databases.  

To predict new DTIs, we employ the CV on all non-interacting pairs which is also utilized in~\cite{Olayan2018DDR:Approaches}. Specifically, all non-interacting pairs are split into ten folds. For each fold of CV, MF2A is trained using all interacting pairs and nine folds of non-interacting pairs and predicts interactivity scores for the one test fold. After accomplishing the ten folds CV, the prediction scores of all non-interacting pairs are obtained. At last, we choose the top 5 ranked non-interacting pairs as the candidate newly discovered interactions of MF2A. 
To further verify the reliability of those new interactions, we check whether there is any evidence supporting them in the latest version of KEGG (K)~ \cite{Kanehisa2017KEGG:Drugs}, DrugBank (DB)~\cite{Wishart2018DrugBank2018}, ChEMBL (C) ~\cite{Mendez2019ChEMBL:Data}, DrugCentral (DC)\cite{Avram2021DrugCentralRepositioning}, and Matador (M)~\cite{Gunther2008SuperTargetRelationships} databases.
Table \ref{tab:newDTIs} lists the top 5 new DTIs found by MF2A on the original golden standard and Luo's dataset, along with the reference databases where the corresponding evidences are found. We can see that 24 predicted new interactions (96\%) are confirmed by at least one external database, demonstrating the reliability of MF2A for novel DTI predictions. 

Moreover, we report the top 5 new DTIs found by training MF2A on the four updated golden standard datasets in the supplementary Table S4. Those novel DTIs, which are not recorded in current databases, are good candidates for validation \textit{in vitro} experiments to check their biological efficiency and clinical meaning.




\begin{table}[h]
\centering
\setlength{\tabcolsep}{1pt}
\caption{Top 5 new DTIs found by MF2A from original golden standard and Luo's datasets}
\label{tab:newDTIs}
\begin{tabular}{@{}ccccccc@{}}
\toprule
Dataset & Drug ID & Drug name & Target ID & Target name & Score & Database \\ \midrule
\multirow{5}{*}{NR} & D00094 & Tretinoin & hsa6095 & RORA & 0.974 & DC \\
 & D00506 & Phenobarbital & hsa8856 & NR1I2 & 0.962 & DB, M, DC \\
 & D00316 & Etretinate & hsa6096 & RORB & 0.96 & -\\
 & D00443 & Spironolactone & hsa5241 & PGR & 0.958 & DB, DC \\
 & D00163 & Chenodiol & hsa7421 & VDR & 0.957 & C \\ \hline
\multirow{5}{*}{GPCR} & D04625 & Isoetharine & hsa154 & ADRB2 & 0.62 & K, DB \\
 & D00559 &\begin{tabular}[c]{@{}c@{}}Pramipexole \\ dihydrochloride\end{tabular} & hsa1813 & DRD2 & 0.619 & K, DB \\
 & D00503 & Perphenazine & hsa3356 & HTR2A & 0.615 & DC \\
 & D02671 & Mesoridazine & hsa3269 & HRH1 & 0.614 & DC \\
 & D02357 & Methysergide & hsa1813 & DRD2 & 0.614 & DC \\ \hline
\multirow{5}{*}{IC} & D00538 & Zonisamide & hsa6331 & SCN5A & 0.636 & K, DB \\
 & D00438 & Nimodipine & hsa779 & CACNA1S & 0.635 & K, DB, DC \\
 & D00552 & Benzocaine & hsa6331 & SCN5A & 0.627 & K, DC \\
 & D03365 & Nicotine & hsa1137 & CHRNA4 & 0.624 & K, DC \\
 & D02347 & Dantrolene & hsa6263 & RYR3 & 0.623 & DC \\ \hline
\multirow{5}{*}{E} & D00528 & Caffeine & hsa1549 & CYP2A7 & 0.814 & M \\
 & D00139 & Methoxsalen & hsa1543 & CYP1A1 & 0.798 & DB, M \\
 & D00437 & Nifedipine & hsa1559 & CYP2C9 & 0.775 & DB, C, M \\
 & D00097 & Salicylic acid & hsa5743 & PTGS2 & 0.769 & DB, M \\
 & D00043 & Isoflurophate & hsa1991 & ELANE & 0.76 & M \\ \hline
\multirow{5}{*}{Luo} & DB00543 & Amoxapine & P28223 & HTR2A & 0.734 & DB \\
 & DB00929 & Misoprostol & P34995 & PTGER1 & 0.733 & DB \\
 & DB00696 & Ergotamine & P08908 & HTR1A & 0.729 & DB \\
 & DB06216 & Asenapine & P28221 & HTR1D & 0.728 & DC \\
 & DB06216 & Asenapine & P21918 & DRD5 & 0.726 & DB \\ \bottomrule
\end{tabular}
\end{table}

\section{Conclusion}
This work proposed a similarity integration method, LIC, which leverages the local interaction consistency based linear weighting to generate the combined similarity that enhances the utility of more reliable views.
We further developed three MF based DTI prediction approaches, namely MFAUPR, MFAUC and MF2A, that optimize AUPR, AUC and both of these metrics, respectively. 
The three MF methods are able to handle multiple similarities input by incorporating LIC as a pre-processing step.
Empirical studies demonstrate that MFAUPR and MFAUC outperform the state of the art in terms of the metric they optimize, and MF2A not only achieves the best results in terms of both AUPR and AUC, but also successfully predicts trustworthy new DTIs that are not reported in the original datasets. In addition, the superiority of LIC compared to the prevalent similarity integration methods is verified as well. 

Current DTI datasets typically contain many drug-target pairs that actually interact, but have not yet been detected due to the complex and costly experimental verification process. Simply treating these missing DTIs as non-interacting ones would lead to crucial information loss and the existence of noise in the data. In the future, we plan to extend our methods to handle missing interactions properly by incorporating a data imputation strategy.

\ifCLASSOPTIONcompsoc
  \section*{Acknowledgments}
\else
  \section*{Acknowledgment}
\fi

Bin Liu was supported from the China Scholarship Council (CSC) under the Grant CSC No.201708500095.

\bibliographystyle{IEEEtran} %
\bibliography{LB.bib}

\begin{thebibliography}{10}
\providecommand{\url}[1]{#1}
\csname url@samestyle\endcsname
\providecommand{\newblock}{\relax}
\providecommand{\bibinfo}[2]{#2}
\providecommand{\BIBentrySTDinterwordspacing}{\spaceskip=0pt\relax}
\providecommand{\BIBentryALTinterwordstretchfactor}{4}
\providecommand{\BIBentryALTinterwordspacing}{\spaceskip=\fontdimen2\font plus
\BIBentryALTinterwordstretchfactor\fontdimen3\font minus
  \fontdimen4\font\relax}
\providecommand{\BIBforeignlanguage}[2]{{%
\expandafter\ifx\csname l@#1\endcsname\relax
\typeout{** WARNING: IEEEtran.bst: No hyphenation pattern has been}%
\typeout{** loaded for the language `#1'. Using the pattern for}%
\typeout{** the default language instead.}%
\else
\language=\csname l@#1\endcsname
\fi
#2}}
\providecommand{\BIBdecl}{\relax}
\BIBdecl

\bibitem{Bagherian2021MachinePaper}
M.~Bagherian, E.~Sabeti, K.~Wang, M.~A. Sartor, Z.~Nikolovska-Coleska, and
  K.~Najarian, ``{Machine learning approaches and databases for prediction of
  drug-target interaction: A survey paper},'' \emph{Briefings in
  Bioinformatics}, vol.~22, no.~1, pp. 247--269, 2021.

\bibitem{Jacob2008Protein-ligandApproach}
L.~Jacob and J.~P. Vert, ``{Protein-ligand interaction prediction: An improved
  chemogenomics approach},'' \emph{Bioinformatics}, vol.~24, no.~19, pp.
  2149--2156, 2008.

\bibitem{Opella2013StructureSpectroscopy}
S.~J. Opella, ``{Structure determination of membrane proteins by nuclear
  magnetic resonance spectroscopy},'' \emph{Annual Review of Analytical
  Chemistry}, vol.~6, no.~1, pp. 305--328, 2013.

\bibitem{Ezzat2018ComputationalSurvey}
A.~Ezzat, M.~Wu, X.~L. Li, and C.~K. Kwoh, ``{Computational prediction of
  drug-target interactions using chemogenomic approaches: an empirical
  survey},'' \emph{Briefings in Bioinformatics}, vol.~20, no.~4, pp.
  1337--1357, 2018.

\bibitem{Ezzat2017Drug-targetFactorization}
A.~Ezzat, P.~Zhao, M.~Wu, X.~L. Li, and C.~K. Kwoh, ``{Drug-target interaction
  prediction with graph regularized matrix factorization},'' \emph{IEEE/ACM
  Transactions on Computational Biology and Bioinformatics}, vol.~14, no.~3,
  pp. 646--656, 2017.

\bibitem{Airola2018FastTrick}
A.~Airola and T.~Pahikkala, ``{Fast kronecker product kernel methods via
  generalized vec trick},'' \emph{IEEE Transactions on Neural Networks and
  Learning Systems}, vol.~29, no.~8, pp. 3374--3387, 2018.

\bibitem{Olayan2018DDR:Approaches}
R.~S. Olayan, H.~Ashoor, and V.~B. Bajic, ``{DDR: Efficient computational
  method to predict drug-Target interactions using graph mining and machine
  learning approaches},'' \emph{Bioinformatics}, vol.~7, no.~34, pp.
  1164--1173, 2018.

\bibitem{Xuan2021IntegratingPrediction}
P.~Xuan, Y.~Zhang, H.~Cui, T.~Zhang, M.~Guo, and T.~Nakaguchi, ``{Integrating
  multi-scale neighbouring topologies and cross-modal similarities for
  drug–protein interaction prediction},'' \emph{Briefings in Bioinformatics},
  vol.~22, no.~5, p. bbab119, 2021.

\bibitem{Nascimento2016APrediction}
A.~C. Nascimento, R.~B. Prud{\^{e}}ncio, and I.~G. Costa, ``{A multiple kernel
  learning algorithm for drug-target interaction prediction},'' \emph{BMC
  Bioinformatics}, vol.~17, no.~1, pp. 1--16, 2016.

\bibitem{Wan2019NeoDTI:Interactions}
F.~Wan, L.~Hong, A.~Xiao, T.~Jiang, and J.~Zeng, ``{NeoDTI: Neural integration
  of neighbor information from a heterogeneous network for discovering new
  drug-target interactions},'' \emph{Bioinformatics}, vol.~35, no.~1, pp.
  104--111, 2019.

\bibitem{Ding2020IdentificationFusion}
Y.~Ding, J.~Tang, and F.~Guo, ``{Identification of drug–target interactions
  via dual laplacian regularized least squares with multiple kernel fusion},''
  \emph{Knowledge-Based Systems}, vol. 204, p. 106254, 2020.

\bibitem{Schrynemackers2013OnNetworks}
M.~Schrynemackers, R.~K{\"{u}}ffner, and P.~Geurts, ``{On protocols and
  measures for the validation of supervised methods for the inference of
  biological networks},'' \emph{Frontiers in Genetics}, vol.~4, p. 262, 2013.

\bibitem{Davis2006TheCurves}
J.~Davis and M.~Goadrich, ``{The relationship between Precision-Recall and ROC
  curves},'' in \emph{International Conference on Machine Learning}, 2006, pp.
  233--240.

\bibitem{Pliakos2020Drug-targetReconstruction}
K.~Pliakos and C.~Vens, ``{Drug-target interaction prediction with
  tree-ensemble learning and output space reconstruction},'' \emph{BMC
  Bioinformatics}, vol.~21, no.~1, pp. 1--11, 2020.

\bibitem{Golkov2020DeepFunctions}
V.~Golkov, A.~Becker, D.~T. Plop, D.~{\v{C}}uturilo, N.~Davoudi, J.~Mendenhall,
  R.~Moretti, J.~Meiler, and D.~Cremers, ``{Deep learning for virtual
  screening: Five reasons to use ROC cost functions},'' \emph{arXiv}, 2020.

\bibitem{Peska2017Drug-targetApproach}
L.~Peska, K.~Buza, and J.~Koller, ``{Drug-target interaction prediction: A
  Bayesian ranking approach},'' \emph{Computer Methods and Programs in
  Biomedicine}, vol. 152, pp. 15--21, 2017.

\bibitem{Chen2018MachinePrediction}
R.~Chen, X.~Liu, S.~Jin, J.~Lin, and J.~Liu, ``{Machine learning for
  drug-target interaction prediction},'' \emph{Molecules}, vol.~23, no.~9, p.
  2208, 2018.

\bibitem{Zheng2013CollaborativeInteractions}
X.~Zheng, H.~Ding, H.~Mamitsuka, and S.~Zhu, ``{Collaborative matrix
  factorization with multiple similarities for predicting drug-Target
  interactions},'' in \emph{ACM International Conference on Knowledge Discovery
  and Data Mining}, 2013, pp. 1025--1033.

\bibitem{Liu2016NeighborhoodPrediction}
Y.~Liu, M.~Wu, C.~Miao, P.~Zhao, and X.~L. Li, ``{Neighborhood regularized
  logistic matrix factorization for drug-target interaction prediction},''
  \emph{PLoS Computational Biology}, vol.~12, no.~2, p. e1004760, 2016.

\bibitem{Shi2012TFMAP:Recommendation}
Y.~Shi, A.~Karatzoglou, L.~Baltrunas, M.~Larson, A.~Hanjalic, and N.~Oliver,
  ``{TFMAP: Optimizing MAP for top-n context-aware recommendation},'' in
  \emph{International Conference on Research and Development in Information
  Retrieval}, 2012, pp. 155--164.

\bibitem{dhanjal2015auc}
C.~Dhanjal, R.~Gaudel, and S.~Clemencon, ``{AUC optimisation and collaborative
  filtering},'' \emph{arXiv}, 2015.

\bibitem{Pahikkala2015TowardPredictions}
T.~Pahikkala, A.~Airola, S.~Pietil{\"{a}}, S.~Shakyawar, A.~Szwajda, J.~Tang,
  and T.~Aittokallio, ``{Toward more realistic drug-target interaction
  predictions},'' \emph{Briefings in Bioinformatics}, vol.~16, no.~2, pp.
  325--337, 2015.

\bibitem{Yamanishi2008PredictionSpaces}
Y.~Yamanishi, M.~Araki, A.~Gutteridge, W.~Honda, and M.~Kanehisa, ``{Prediction
  of drug-target interaction networks from the integration of chemical and
  genomic spaces},'' \emph{Bioinformatics}, vol.~24, no.~13, p. i232–i240,
  2008.

\bibitem{Shi2015PredictingClustering}
J.~Y. Shi, S.~M. Yiu, Y.~Li, H.~C. Leung, and F.~Y. Chin, ``{Predicting
  drug-target interaction for new drugs using enhanced similarity measures and
  super-target clustering},'' \emph{Methods}, vol.~83, pp. 98--104, 2015.

\bibitem{Liu2021Drug-TargetRecovery}
B.~Liu, K.~Pliakos, C.~Vens, and G.~Tsoumakas, ``{Drug-target interaction
  prediction via an ensemble of weighted nearest neighbors with interaction
  recovery},'' \emph{Applied Intelligence}, vol. Available, 2021.

\bibitem{Mei2013Drug-targetNeighbors}
J.~P. Mei, C.~K. Kwoh, P.~Yang, X.~L. Li, and J.~Zheng, ``{Drug-target
  interaction prediction by learning from local information and neighbors},''
  \emph{Bioinformatics}, vol.~29, no.~2, pp. 238--245, 2013.

\bibitem{vanLaarhoven2013PredictingProfile}
T.~van Laarhoven and E.~Marchiori, ``{Predicting drug-target interactions for
  new drug compounds using a weighted nearest neighbor profile},'' \emph{PLoS
  ONE}, vol.~8, no.~6, p. e66952, 2013.

\bibitem{Li2019DrugEmbedding}
L.~Li and M.~Cai, ``{Drug target prediction by multi-view low rank
  embedding},'' \emph{IEEE/ACM Transactions on Computational Biology and
  Bioinformatics}, vol.~16, no.~5, pp. 1712--1721, 2019.

\bibitem{Wang2013DrugInference}
W.~Wang, S.~Yang, and L.~Jing, ``{Drug target predictions based on
  heterogeneous graph inference},'' in \emph{Pacific Symposium on
  Biocomputing}, 2013, pp. 53--64.

\bibitem{Chen2012Drug-targetNetwork}
X.~Chen, M.~X. Liu, and G.~Y. Yan, ``{Drug-target interaction prediction by
  random walk on the heterogeneous network},'' \emph{Molecular BioSystems},
  vol.~8, no.~7, pp. 1970--1978, 2012.

\bibitem{Chu2021DTI-CDF:Features}
Y.~Chu, A.~C. Kaushik, X.~Wang, W.~Wang, Y.~Zhang, X.~Shan, D.~R. Salahub,
  Y.~Xiong, and D.-Q. Wei, ``{DTI-CDF: a cascade deep forest model towards the
  prediction of drug-target interactions based on hybrid features},''
  \emph{Briefings in Bioinformatics}, vol.~22, no.~1, pp. 451--462, 2021.

\bibitem{Thafar2020DTiGEMS+:Techniques}
M.~A. Thafar, M.~A. Thafar, R.~S. Olayan, R.~S. Olayan, H.~Ashoor, H.~Ashoor,
  S.~Albaradei, S.~Albaradei, V.~B. Bajic, X.~Gao, T.~Gojobori, T.~Gojobori,
  and M.~Essack, ``{DTiGEMS+: Drug-target interaction prediction using graph
  embedding, graph mining, and similarity-based techniques},'' \emph{Journal of
  Cheminformatics}, vol.~12, no.~1, pp. 1--17, 2020.

\bibitem{Luo2017AInformation}
Y.~Luo, X.~Zhao, J.~Zhou, J.~Yang, Y.~Zhang, W.~Kuang, J.~Peng, L.~Chen, and
  J.~Zeng, ``{A network integration approach for drug-target interaction
  prediction and computational drug repositioning from heterogeneous
  information},'' \emph{Nature Communications}, vol.~8, no.~1, pp. 1--13, 2017.

\bibitem{Sun2020GraphInteractions}
C.~Sun, P.~Xuan, T.~Zhang, and Y.~Ye, ``{Graph convolutional autoencoder and
  generative adversarial network-based method for predicting drug-target
  interactions},'' \emph{IEEE/ACM Transactions on Computational Biology and
  Bioinformatics}, 2020.

\bibitem{Pliakos2021PredictingPartitioning}
K.~Pliakos, C.~Vens, and G.~Tsoumakas, ``{Predicting drug-target interactions
  with multi-label classification and label partitioning},'' \emph{IEEE/ACM
  Transactions on Computational Biology and Bioinformatics}, vol.~18, no.~4,
  pp. 1596--1607, 2021.

\bibitem{Sachdev2019APrediction}
K.~Sachdev and M.~K. Gupta, ``{A comprehensive review of feature based methods
  for drug target interaction prediction},'' \emph{Journal of Biomedical
  Informatics}, vol.~93, p. 103159, 2019.

\bibitem{Gao2018InterpretableRepresentation}
K.~Y. Gao, A.~Fokoue, H.~Luo, A.~Iyengar, S.~Dey, and P.~Zhang,
  ``{Interpretable drug target prediction using deep neural representation},''
  in \emph{International Joint Conference on Artificial Intelligence}, 2018,
  pp. 3371--3377.

\bibitem{Yu2021KenDTI:Prediction}
Z.~Yu, J.~Lu, Y.~Jin, and Y.~Yang, ``{KenDTI: an ensemble model based on
  network integration and CNN for drug-target interaction prediction},''
  \emph{IEEE/ACM Transactions on Computational Biology and Bioinformatics},
  vol.~18, no.~4, pp. 1305--1314, 2021.

\bibitem{liu2021multi}
B.~Liu, K.~Blekas, and G.~Tsoumakas, ``{Multi-Label Sampling based on Local
  Label Imbalance},'' \emph{Pattern Recognition}, vol. 122, p. 108294, 2021.

\bibitem{Liu2018RegularizedSurvey}
J.~X. Liu, D.~Wang, Y.~L. Gao, C.~H. Zheng, Y.~Xu, and J.~Yu, ``{Regularized
  non-negative matrix factorization for identifying differentially expressed
  genes and clustering samples: A survey},'' \emph{IEEE/ACM Transactions on
  Computational Biology and Bioinformatics}, vol.~15, no.~3, pp. 974--987,
  2018.

\bibitem{cambridge2009online}
C.~D. Manning, P.~Raghavan, and H.~Sch{\"{u}}tze, \emph{{An introduction to
  information retrieval}}.\hskip 1em plus 0.5em minus 0.4em\relax Cambridge
  University Press, 2009.

\bibitem{He2018HashingRank}
K.~He, F.~Cakir, S.~A. Bargal, and S.~Sclaroff, ``{Hashing as tie-aware
  learning to rank},'' in \emph{IEEE Conference on Computer Vision and Pattern
  Recognition}, 2018, pp. 4023--4032.

\bibitem{Revaud2019LearningLoss}
J.~Revaud, J.~Almazan, R.~Rezende, and C.~D. Souza, ``{Learning with average
  precision: Training image retrieval with a listwise loss},'' in \emph{IEEE
  International Conference on Computer Vision}, 2019, pp. 5106--5115.

\bibitem{Gao2015OnOptimization}
W.~Gao and Z.~H. Zhou, ``{On the consistency of AUC pairwise optimization},''
  in \emph{International Joint Conference on Artificial Intelligence}, 2015,
  pp. 939--945.

\bibitem{Gultekin2020MBA:Optimization}
S.~Gultekin, A.~Saha, A.~Ratnaparkhi, and J.~Paisley, ``{MBA: Mini-batch AUC
  optimization},'' \emph{IEEE Transactions on Neural Networks and Learning
  Systems}, vol.~31, no.~12, pp. 5561--5574, 2020.

\bibitem{Duchi2011AdaptiveOptimization}
J.~Duchi, E.~Hazan, and Y.~Singer, ``{Adaptive subgradient methods for online
  learning and stochastic optimization},'' \emph{Journal of Machine Learning
  Research}, vol.~12, pp. 2121--2159, 2011.

\bibitem{Kanehisa2017KEGG:Drugs}
M.~Kanehisa, M.~Furumichi, M.~Tanabe, Y.~Sato, and K.~Morishima, ``{KEGG: New
  perspectives on genomes, pathways, diseases and drugs},'' \emph{Nucleic Acids
  Research}, vol.~45, no.~D1, pp. D353--D361, 2017.

\bibitem{Wishart2018DrugBank2018}
D.~S. Wishart, Y.~D. Feunang, A.~C. Guo, E.~J. Lo, A.~Marcu, J.~R. Grant,
  T.~Sajed, D.~Johnson, C.~Li, Z.~Sayeeda, N.~Assempour, I.~Iynkkaran, Y.~Liu,
  A.~MacIejewski, N.~Gale, A.~Wilson, L.~Chin, R.~Cummings, D.~Le, A.~Pon,
  C.~Knox, and M.~Wilson, ``{DrugBank 5.0: A major update to the DrugBank
  database for 2018},'' \emph{Nucleic Acids Research}, vol.~46, no.~D1, pp.
  D1074--D1082, 2018.

\bibitem{Mendez2019ChEMBL:Data}
D.~Mendez, A.~Gaulton, A.~P. Bento, J.~Chambers, M.~De~Veij, E.~F{\'{e}}lix,
  M.~P. Magari{\~{n}}os, J.~F. Mosquera, P.~Mutowo, M.~Nowotka,
  M.~Gordillo-Mara{\~{n}}{\'{o}}n, F.~Hunter, L.~Junco, G.~Mugumbate,
  M.~Rodriguez-Lopez, F.~Atkinson, N.~Bosc, C.~J. Radoux, A.~Segura-Cabrera,
  A.~Hersey, and A.~R. Leach, ``{ChEMBL: Towards direct deposition of bioassay
  data},'' \emph{Nucleic Acids Research}, vol.~47, no.~D1, pp. D930--D940,
  2019.

\bibitem{Hattori2003DevelopmentPathways}
M.~Hattori, Y.~Okuno, S.~Goto, and M.~Kanehisa, ``{Development of a chemical
  structure comparison method for integrated analysis of chemical and genomic
  information in the metabolic pathways},'' \emph{Journal of the American
  Chemical Society}, vol. 125, no.~39, pp. 11\,853--11\,865, 2003.

\bibitem{Takarabe2012DrugApproach}
M.~Takarabe, M.~Kotera, Y.~Nishimura, S.~Goto, and Y.~Yamanishi, ``{Drug target
  prediction using adverse event report systems: a pharmacogenomic approach},''
  \emph{Bioinformatics}, vol.~28, no.~18, p. i611–i618, 2012.

\bibitem{Kuhn2016TheEffects}
M.~Kuhn, I.~Letunic, L.~J. Jensen, and P.~Bork, ``{The SIDER database of drugs
  and side effects},'' \emph{Nucleic Acids Research}, vol.~44, no.~D1, pp.
  D1075--D1079, 2016.

\bibitem{Smith1981IdentificationSubsequences}
T.~F. Smith and M.~S. Waterman, ``{Identification of common molecular
  subsequences},'' \emph{Journal of Molecular Biology}, vol. 147, no.~1, pp.
  195--197, 1981.

\bibitem{Leslie2002TheClassification}
C.~Leslie, E.~Eskin, and W.~S. Noble, ``{The spectrum kernel: a string kernel
  for SVM protein classification},'' \emph{Pacific Symposium on Biocomputing},
  pp. 564--575, 2002.

\bibitem{Barrell2009TheResource}
D.~Barrell, E.~Dimmer, R.~P. Huntley, D.~Binns, C.~O'Donovan, and R.~Apweiler,
  ``{The GOA database in 2009 - an integrated gene ontology annotation
  resource},'' \emph{Nucleic Acids Research}, vol.~37, no. SUPPL. 1, pp.
  396--403, 2009.

\bibitem{KeshavaPrasad2009HumanUpdate}
T.~S. Keshava~Prasad, R.~Goel, K.~Kandasamy, S.~Keerthikumar, S.~Kumar,
  S.~Mathivanan, D.~Telikicherla, R.~Raju, B.~Shafreen, A.~Venugopal,
  L.~Balakrishnan, A.~Marimuthu, S.~Banerjee, D.~S. Somanathan, A.~Sebastian,
  S.~Rani, S.~Ray, C.~J. Harrys~Kishore, S.~Kanth, M.~Ahmed, M.~K. Kashyap,
  R.~Mohmood, Y.~I. Ramachandra, V.~Krishna, B.~A. Rahiman, S.~Mohan,
  P.~Ranganathan, S.~Ramabadran, R.~Chaerkady, and A.~Pandey, ``{Human protein
  reference database-2009 update},'' \emph{Nucleic Acids Research}, vol.~37,
  no. suppl{\_}1, p. D767–D772, 2009.

\bibitem{Davis2013The2013}
A.~P. Davis, C.~G. Murphy, R.~Johnson, J.~M. Lay, K.~Lennon-Hopkins,
  C.~Saraceni-Richards, D.~Sciaky, B.~L. King, M.~C. Rosenstein, T.~C. Wiegers,
  and C.~J. Mattingly, ``{The comparative toxicogenomics database: Update
  2013},'' \emph{Nucleic Acids Research}, vol.~41, no.~D1, pp. D1104--D1114,
  2013.

\bibitem{Benavoli2016}
A.~Benavoli, G.~Corani, and F.~Mangili, ``{Should we Really Use Post-Hoc Tests
  Based on Mean-Ranks?}'' \emph{Journal of Machine Learning Research}, vol.~17,
  no.~1, pp. 1--10, 2016.

\bibitem{Avram2021DrugCentralRepositioning}
S.~Avram, C.~G. Bologa, J.~Holmes, G.~Bocci, T.~B. Wilson, D.~T. Nguyen,
  R.~Curpan, L.~Halip, A.~Bora, J.~J. Yang, J.~Knockel, S.~Sirimulla, O.~Ursu,
  and T.~I. Oprea, ``{DrugCentral 2021 supports drug discovery and
  repositioning},'' \emph{Nucleic Acids Research}, vol.~49, no.~D1, pp.
  D1160--D1169, 2021.

\bibitem{Gunther2008SuperTargetRelationships}
S.~G{\"{u}}nther, M.~Kuhn, M.~Dunkel, M.~Campillos, C.~Senger, E.~Petsalaki,
  J.~Ahmed, E.~G. Urdiales, A.~Gewiess, L.~J. Jensen, R.~Schneider, R.~Skoblo,
  R.~B. Russell, P.~E. Bourne, P.~Bork, and R.~Preissner, ``{SuperTarget and
  Matador: Resources for exploring drug-target relationships},'' \emph{Nucleic
  Acids Research}, vol.~36, no. SUPPL. 1, pp. D919--D922, 2008.

\end{thebibliography}


\begin{thebibliography}{1}
\providecommand{\url}[1]{#1}
\csname url@samestyle\endcsname
\providecommand{\newblock}{\relax}
\providecommand{\bibinfo}[2]{#2}
\providecommand{\BIBentrySTDinterwordspacing}{\spaceskip=0pt\relax}
\providecommand{\BIBentryALTinterwordstretchfactor}{4}
\providecommand{\BIBentryALTinterwordspacing}{\spaceskip=\fontdimen2\font plus
\BIBentryALTinterwordstretchfactor\fontdimen3\font minus
  \fontdimen4\font\relax}
\providecommand{\BIBforeignlanguage}[2]{{%
\expandafter\ifx\csname l@#1\endcsname\relax
\typeout{** WARNING: IEEEtran.bst: No hyphenation pattern has been}%
\typeout{** loaded for the language `#1'. Using the pattern for}%
\typeout{** the default language instead.}%
\else
\language=\csname l@#1\endcsname
\fi
#2}}
\providecommand{\BIBdecl}{\relax}
\BIBdecl

\bibitem{Nascimento2016APrediction}
A.~C. Nascimento, R.~B. Prud{\^{e}}ncio, and I.~G. Costa, ``{A multiple kernel
  learning algorithm for drug-target interaction prediction},'' \emph{BMC
  Bioinformatics}, vol.~17, no.~1, pp. 1--16, 2016.

\bibitem{vanLaarhoven2011GaussianInteraction}
T.~van Laarhoven, S.~B. Nabuurs, and E.~Marchiori, ``{Gaussian interaction
  profile kernels for predicting drug-target interaction},''
  \emph{Bioinformatics}, vol.~27, no.~21, pp. 3036--3043, 2011.

\bibitem{Mei2013Drug-targetNeighbors}
J.~P. Mei, C.~K. Kwoh, P.~Yang, X.~L. Li, and J.~Zheng, ``{Drug-target
  interaction prediction by learning from local information and neighbors},''
  \emph{Bioinformatics}, vol.~29, no.~2, pp. 238--245, 2013.

\end{thebibliography}

\ifCLASSOPTIONcaptionsoff
  \newpage
\fi



%

\begin{IEEEbiography}[{\includegraphics[width=1in,height=1.25in,clip,keepaspectratio]{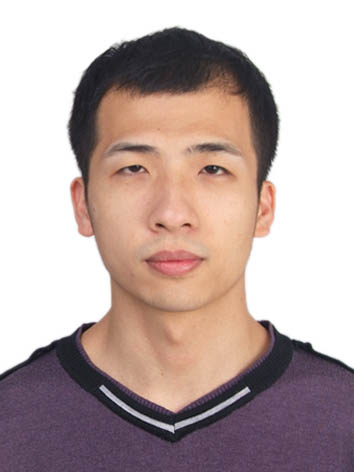}}]{Bin Liu}
received an M.S. degree in computer science from Chongqing University of Posts and Telecommunications, China in 2016. He is currently pursuing a Ph.D. degree in computer science from Aristotle University of Thessaloniki, Greece. His research interests include multi-label learning, class imbalance and bioinformatics.
\end{IEEEbiography}

\begin{IEEEbiography}[{\includegraphics[width=1in,height=1.25in,clip,keepaspectratio]{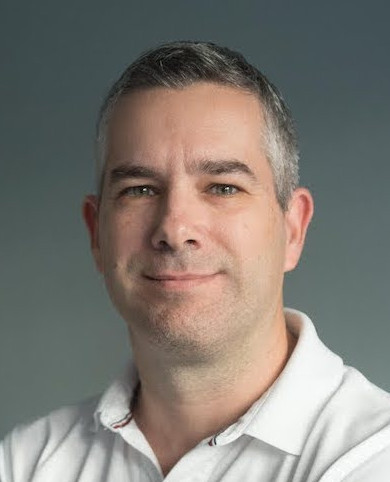}}]{Grigorios Tsoumakas} is an Associate Professor of Machine Learning and Knowledge Discovery at the School of Informatics of the Aristotle University of Thessaloniki (AUTH) in Greece. He received a degree in Computer Science from AUTH in 1999, an MSc in Artificial Intelligence from the University of Edinburgh, United Kingdom, in 2000 and a PhD in Computer Science from AUTH in 2005. His research expertise focuses on supervised learning techniques (ensemble methods, multi-target prediction) and natural language processing (semantic indexing, keyphrase extraction, summarization). He has published more than 100 research papers and according to Google Scholar he has more than 14,000 citations and an h-index of 43. Dr. Tsoumakas is a senior member of the ACM and an action editor of the Data Mining and Knowledge Discovery journal. His honors include receiving the European Conference on Machine Learning and Principles and Practice of Knowledge Discovery in Databases (ECML PKDD) 10-Year Test of Time Award in 2017. He is an advocate of applied research that matters and has worked as a machine learning engineer, researcher and consultant in several national, international and private sector funded R\&D projects. In February 2019 he co-founded Medoid AI, a spin-off company of the Aristotle University of Thessaloniki developing AI products and custom solutions based on machine learning technology. 
\end{IEEEbiography}




\end{document}


%
\title{Optimizing Area Under the Curve Measures via Matrix Factorization for Predicting Drug-Target Interaction with Multiple Similarities}
%
%
%
%

\author{Bin~Liu and Grigorios~Tsoumakas
\IEEEcompsocitemizethanks{\IEEEcompsocthanksitem B. Liu and G. Tsoumakas are with the School of Informatics, Aristotle University of Thessaloniki, Thessaloniki 54124, Greece. \protect\\  E-mail: \{binliu, greg\}@csd.auth.gr} 
}

%
%

\markboth{IEEE/ACM Transactions on Computational Biology and Bioinformatics}%
{Shell \MakeLowercase{\textit{et al.}}: Bare Demo of IEEEtran.cls for Computer Society Journals}
\maketitle

\IEEEdisplaynontitleabstractindextext

%
\IEEEpeerreviewmaketitle

\renewcommand{\thesection}{S\arabic{section}}
\renewcommand{\thetable}{S\arabic{table}}
\renewcommand{\thefigure}{S\arabic{figure}}

\section{Similarities for Golden Standard Datasets}

The information and local interaction consistency of similarities from~\cite{Nascimento2016APrediction} are shown in Table \ref{Stab:similarities}, where the similarities with the higher local interaction consistency, which are highlighted in bold font, are used in our study. 
For Gaussian interaction profile (GIP) based drug and target similarities that are not included in the Table \ref{Stab:similarities}, we will discuss them separately due to their special source information, i.e. the interaction profile.

The GIP similarity is computed based on bipartite network topology information derived from the interaction matrix. Let the interaction profile of a drug $d_i$ (target $t_i$) be $i$-th row ($j$-th column) of the interaction matrix, and the GIP similarity between two drugs (targets) is defined as the Gaussian kernel of their interaction profiles~\cite{vanLaarhoven2011GaussianInteraction}. For simplicity, we use GIP-d and GIP-t to denote GIP similarity for drugs and targets, respectively.
However, the interaction profile of a new drug (target) is a zero-valued vector, as no interactions are known for it. This leads to the reliability of GIP similarity decreases for new entities. Although some methods compute the GIP similarity for new drugs (targets) using the temporary interaction profiles inferred from the interactions of training neighbours, the estimated interaction profiles inevitably containing some false and untagged interactions are still not as ideal as the ground truth ones. 

To check if there is a large discrepancy between the reliability of GIP-d between training and test drugs, we compute the local interaction consistency of GIP-d for training and test drugs separately. Following the 10-fold cross-validation (CV) on drugs used in the prediction setting 2 (S2), for each test fold, ground truth interactions of test drugs are set as the temporary interaction profiles estimated based on the interacting information of training neighbours~\cite{Mei2013Drug-targetNeighbors}. Then the GIP-d is computed based on the processed interaction matrix. 
Next, we compute local interaction consistency of GIP-d for each drug-target pair and average the local interaction consistency of pairs relating to training and test drugs separately to obtain two measures: $LC_{tr}$ and $LC_{te}$. The higher the $LC_{tr}$ ($LC_{te}$), the more reliability of GIP-d for the training (test) drugs. The averaged $LC_{tr}$ and $LC_{te}$ across 10 folds are computed as the final reliability indicators.
The computation of $LC_{tr}$ and $LC_{te}$ for GIP-t is in the similar way, except for using the 10-fold CV on targets instead of CV on drugs.

\begin{figure}[h]
\centering
\subfloat[GIP-d]{\includegraphics[width=0.23\textwidth]{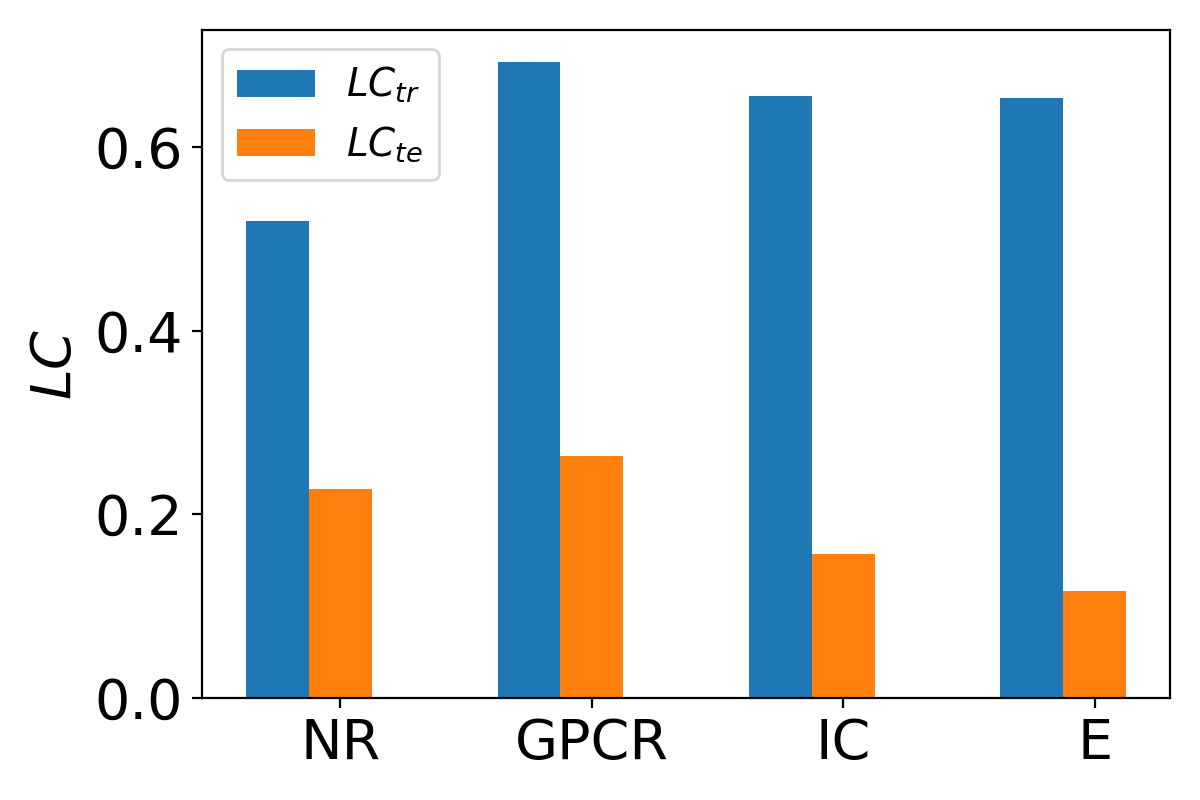}}
\hspace{1em}
\subfloat[GIP-t]{\includegraphics[width=0.23\textwidth]{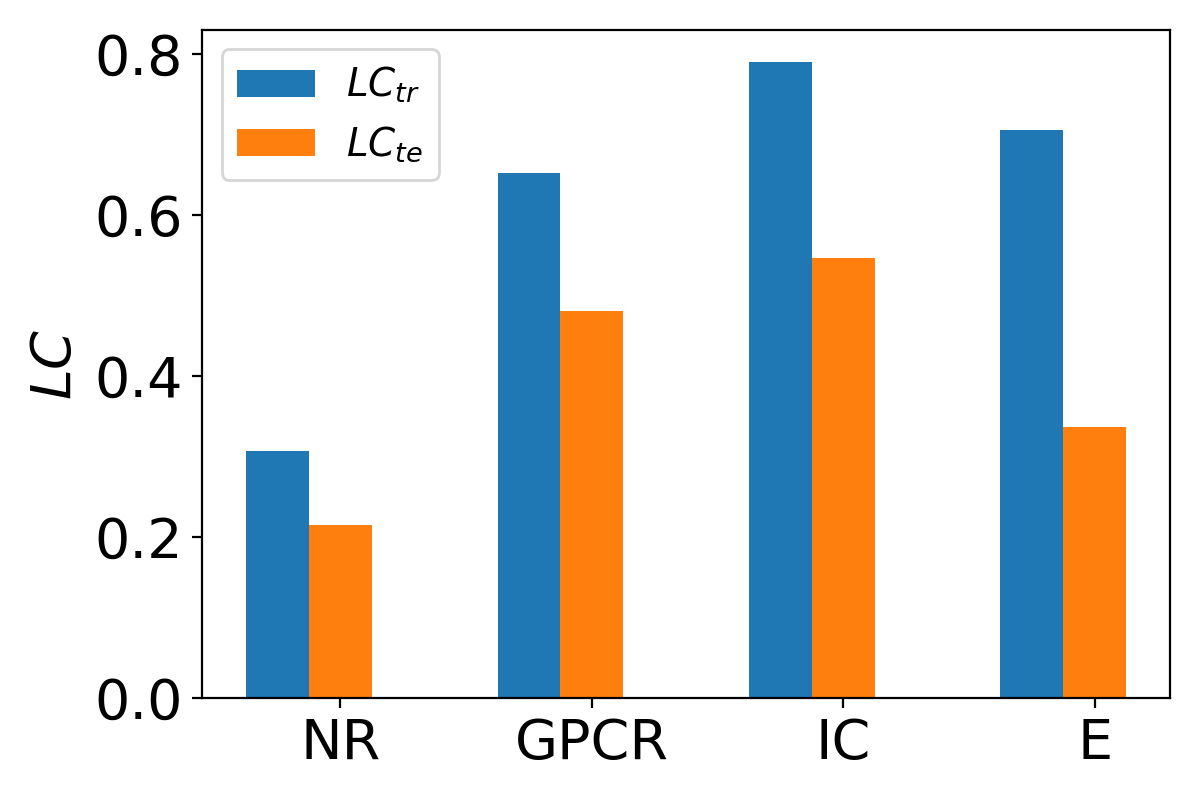}}
\caption{Local interaction consistency of GIP similarities for the training and test set} 
\label{Sfig:GIP_Limb}
\end{figure}

Supplementary Figure \ref{Sfig:GIP_Limb} shows the $LC_{tr}$ and $LC_{te}$ for GIP-d and GIP-t on the four golden standard datasets. In both drug and target aspects, $LC_{tr}$ is much higher than $LC_{te}$, indicating that although the GIP similarity has higher reliability for the training set, its reliability largely decreases for test data. 
This inconsistent reliability of GIP similarity for training and test data has a crucial impact on the proposed local interaction consistency based similarity integration method (LIC). LIC assigns a larger weight to the GIP similarity according to its higher reliability on the training set, which would overestimate the importance of GIP similarity for test data, as the similarity weights of training and test set are identical.

\captionsetup[figure]{name={Supplementary Fig.},labelsep=colon}
\setcounter{table}{0}

Furthermore, we compare the results of MF2A using multiple similarities with and without the GIP similarity. As shown in Table \ref{Stab:GIP_result}, the performance of MF2A deteriorates by adding the GIP similarity in most cases, demonstrating the detriment of the GIP similarity. Therefore, we do not use the GIP similarity in our study.

\captionsetup[table]{name={Supplementary Table},labelsep=colon}
\setcounter{table}{0}

\begin{table*}[!htb]
\centering
\caption{The information and local interaction consistency of similarities from~\cite{Nascimento2016APrediction}}
\label{Stab:similarities}
\begin{tabular}{@{}ccccccccc@{}}
\toprule
& Similarity Name & Computation method & Source & NR & GPCR & IC & E & aveRank \\ \midrule
Drug & \textbf{SIMCOMP} & \textbf{SIMCOMP} & \textbf{Chemical structure} & \textbf{0.42} & \textbf{0.41} & \textbf{0.366} & \textbf{0.302} & \textbf{1} \\
 & \textbf{AERS-bit} & \textbf{Weighted cosine correlation} & \textbf{Side effect} & \textbf{0.322} & \textbf{0.345} & \textbf{0.189} & \textbf{0.155} & \textbf{2.75} \\
 & \textbf{AERS-freq} & \textbf{Weighted cosine correlation} & \textbf{Side effect} & \textbf{0.306} & \textbf{0.354} & \textbf{0.202} & \textbf{0.161} & \textbf{2.25} \\
 & LAMBDA & Lambda-k kernel & Chemical structure & 0.212 & 0.132 & 0.109 & 0.104 & 7 \\
 & MARG & Marginalized Kernel & Chemical structure & 0.203 & 0.147 & 0.093 & 0.093 & 7.5 \\
 & MINMAX & MinMax Kernel & Chemical structure & 0.201 & 0.135 & 0.111 & 0.098 & 7.25 \\
 & \textbf{SIDER} & \textbf{Weighted cosine correlation} & \textbf{Side effect} & \textbf{0.248} & \textbf{0.297} & \textbf{0.159} & \textbf{0.123} & \textbf{4} \\
 & SPEC & Spectrum kernel & Chemical structure & 0.212 & 0.132 & 0.109 & 0.104 & 7 \\
 & TAN & Tanimoto kernel & Chemical structure & 0.196 & 0.141 & 0.126 & 0.114 & 6.25 \\ \hline
Target & \textbf{SW} & \textbf{Normalized Smith–Waterman score} & \textbf{Sequences} & \textbf{0.326} & \textbf{0.573} & \textbf{0.778} & \textbf{0.609} & \textbf{2} \\
 & \textbf{GO} & \textbf{Semantic similarity score} & \textbf{GO annotation} & \textbf{0.225} & \textbf{0.589} & \textbf{0.661} & \textbf{0.564} & \textbf{3.75} \\
 & MIS-k3m1 & Mismatch kernel (k = 3, m = 1) & Sequences & 0.256 & 0.298 & 0.544 & 0.271 & 6.25 \\
 & MIS-k4m1 & Mismatch kernel (k = 4, m = 1) & Sequences & 0.211 & 0.223 & 0.498 & 0.306 & 7.5 \\
 & MIS-k3m2 & Mismatch kernel (k = 3, m = 2) & Sequences & 0.306 & 0.437 & 0.559 & 0.29 & 5 \\
 & MIS-k4m2 & Mismatch kernel (k = 4, m = 2) & Sequences & 0.239 & 0.263 & 0.542 & 0.266 & 7.25 \\
 & PPI & Shortest distance & PPI network & 0.293 & 0.073 & 0.171 & 0.149 & 8 \\
 & \textbf{SPEC-k3} & \textbf{Spectrum kernel (k=3)} & \textbf{Sequences} & \textbf{0.351} & \textbf{0.538} & \textbf{0.639} & \textbf{0.401} & \textbf{3.5} \\
 & \textbf{SPEC-k4} & \textbf{Spectrum kernel (k=4)} & \textbf{Sequences} & \textbf{0.385} & \textbf{0.647} & \textbf{0.762} & \textbf{0.552} & \textbf{1.75}  \\ \bottomrule
\end{tabular}
\end{table*}

\begin{table*}[h]
\centering
\caption{Predicting results of MF2A using the top four similarities with the highest local interaction consistency (Top4) and Top4 similarities plus the GIP similarity (Top4+GIP)}
\label{Stab:GIP_result}
\begin{tabular}{@{}cccc|cc|cc|cc@{}}
\toprule
\multirow{2}{*}{Metric} & Setting & \multicolumn{2}{c|}{S1} & \multicolumn{2}{c|}{S2} & \multicolumn{2}{c|}{S3} & \multicolumn{2}{c}{S4} \\
 & Similarities & Top4 & Top4+GIP & Top4 & Top4+GIP & Top4 & Top4+GIP & Top4 & Top4+GIP \\ \hline
\multirow{4}{*}{AUPR} & NR & \textbf{0.673} & 0.611 & \textbf{0.578} & 0.541 & \textbf{0.588} & 0.567 & 0.289 & \textbf{0.296} \\
 & GPCR & \textbf{0.87} & 0.857 & \textbf{0.551} & 0.457 & \textbf{0.787} & 0.747 & \textbf{0.407} & 0.354 \\
 & IC & \textbf{0.943} & 0.911 & \textbf{0.495} & 0.415 & \textbf{0.863} & 0.858 & \textbf{0.352} & 0.326 \\
 & E & \textbf{0.858} & 0.84 & \textbf{0.422} & 0.383 & \textbf{0.731} & 0.716 & \textbf{0.235} & 0.23 \\ \hline
\multirow{4}{*}{AUC} & NR & \textbf{0.884} & 0.87 & \textbf{0.833} & 0.818 & 0.819 & \textbf{0.82} & 0.649 & \textbf{0.675} \\
 & GPCR & \textbf{0.978} & 0.969 & \textbf{0.924} & 0.884 & \textbf{0.96} & 0.949 & \textbf{0.886} & 0.862 \\
 & IC & \textbf{0.989} & 0.98 & \textbf{0.828} & 0.77 & \textbf{0.967} & 0.964 & \textbf{0.776} & 0.713 \\
 & E & \textbf{0.983} & 0.982 & \textbf{0.892} & 0.825 & \textbf{0.944} & 0.937 & \textbf{0.821} & 0.764 \\  \bottomrule
\end{tabular}
\end{table*}

\section{Supplementary Results}

\begin{table*}[h]
\centering
\caption{Full results of MF2A with difference similarity integration methods in all prediction settings}
\label{Stab:Various_Various_SimI}
\begin{tabular}{ccccccc|ccccc}
\toprule
&  & \multicolumn{5}{c|}{AUPR} & \multicolumn{5}{c}{AUC} \\
 &  & Ave & KA & HISC & SNF & LIC & Ave & KA & HISC & SNF & LIC \\ \hline
\multirow{4}{*}{S1} & NR & 0.655 & 0.665 & 0.665 & 0.611 & \textbf{0.673} & 0.882 & \textbf{0.884} & 0.882 & 0.877 & \textbf{0.884} \\
 & GPCR & 0.868 & 0.869 & \textbf{0.87} & 0.853 & \textbf{0.87} & 0.977 & \textbf{0.978} & 0.977 & 0.972 & \textbf{0.978} \\
 & IC & 0.941 & 0.941 & 0.942 & 0.939 & \textbf{0.943} & 0.988 & \textbf{0.989} & \textbf{0.989} & 0.986 & \textbf{0.989} \\
 & E & 0.857 & \textbf{0.858} & 0.857 & 0.84 & \textbf{0.858} & \textbf{0.984} & \textbf{0.984} & \textbf{0.984} & 0.979 & 0.983 \\
\multicolumn{2}{c}{\textit{AveRank}} & 3.75 & 2.63 & 2.38 & 5 & \textbf{1.25} & 3.25 & \textbf{1.75} & 2.75 & 5 & 2.25 \\ \hline
\multirow{4}{*}{S2} & NR & 0.555 & 0.569 & 0.562 & 0.518 & \textbf{0.578} & 0.832 & \textbf{0.834} & 0.831 & 0.802 & 0.833 \\
 & GPCR & 0.533 & 0.548 & 0.536 & 0.524 & \textbf{0.551} & 0.921 & 0.923 & 0.921 & 0.911 & \textbf{0.924} \\
 & IC & 0.486 & \textbf{0.496} & 0.488 & 0.424 & 0.495 & 0.826 & 0.827 & 0.826 & 0.807 & \textbf{0.828} \\
 & E & 0.392 & 0.406 & 0.395 & 0.37 & \textbf{0.422} & 0.885 & 0.885 & 0.885 & 0.861 & \textbf{0.892} \\
\multicolumn{2}{c}{\textit{AveRank}} & 4 & 1.75 & 3 & 5 & \textbf{1.25} & 3.25 & 2 & 3.5 & 5 & \textbf{1.25} \\ \hline
\multirow{4}{*}{S3} & NR & 0.581 & 0.571 & 0.574 & 0.521 & \textbf{0.588} & \textbf{0.831} & 0.822 & 0.827 & 0.793 & 0.819 \\
 & GPCR & 0.78 & 0.777 & 0.784 & 0.778 & \textbf{0.787} & 0.959 & 0.961 & 0.959 & \textbf{0.962} & 0.96 \\
 & IC & \textbf{0.863} & \textbf{0.863} & \textbf{0.863} & 0.852 & \textbf{0.863} & \textbf{0.967} & \textbf{0.967} & \textbf{0.967} & 0.963 & \textbf{0.967} \\
 & E & 0.731 & \textbf{0.735} & 0.732 & 0.712 & 0.731 & \textbf{0.944} & \textbf{0.944} & \textbf{0.944} & 0.936 & \textbf{0.944} \\
\multicolumn{2}{c}{\textit{AveRank}} & 2.75 & 3.13 & 2.38 & 4.75 & \textbf{2} & 2.63 & \textbf{2.5} & 2.88 & 4 & 3 \\ \hline
\multirow{4}{*}{S4} & NR & 0.294 & 0.284 & \textbf{0.297} & 0.238 & 0.289 & 0.658 & 0.649 & \textbf{0.66} & 0.622 & 0.649 \\
 & GPCR & 0.4 & 0.401 & 0.401 & 0.379 & \textbf{0.407} & 0.883 & 0.884 & 0.883 & 0.874 & \textbf{0.886} \\
 & IC & 0.34 & 0.348 & 0.341 & 0.282 & \textbf{0.352} & 0.772 & 0.773 & 0.771 & 0.758 & \textbf{0.776} \\
 & E & 0.213 & 0.225 & 0.215 & 0.218 & \textbf{0.235} & 0.813 & 0.817 & 0.816 & 0.802 & \textbf{0.821} \\
\multicolumn{2}{c}{\textit{AveRank}} & 3.75 & 2.63 & 2.63 & 4.5 & \textbf{1.5} & 3.13 & 2.38 & 2.88 & 5 & \textbf{1.63} \\ \hline
\multicolumn{2}{c}{\textit{Summary}} & 3.56 & 2.54 & 2.6 & 4.81 & \textbf{1.5} & 3.07 & 2.16 & 3 & 4.75 & \textbf{2.03}  \\ \bottomrule
\end{tabular}
\end{table*}

\begin{figure}[h]
\centering
\subfloat[AUPR]{\includegraphics[width=0.23\textwidth]{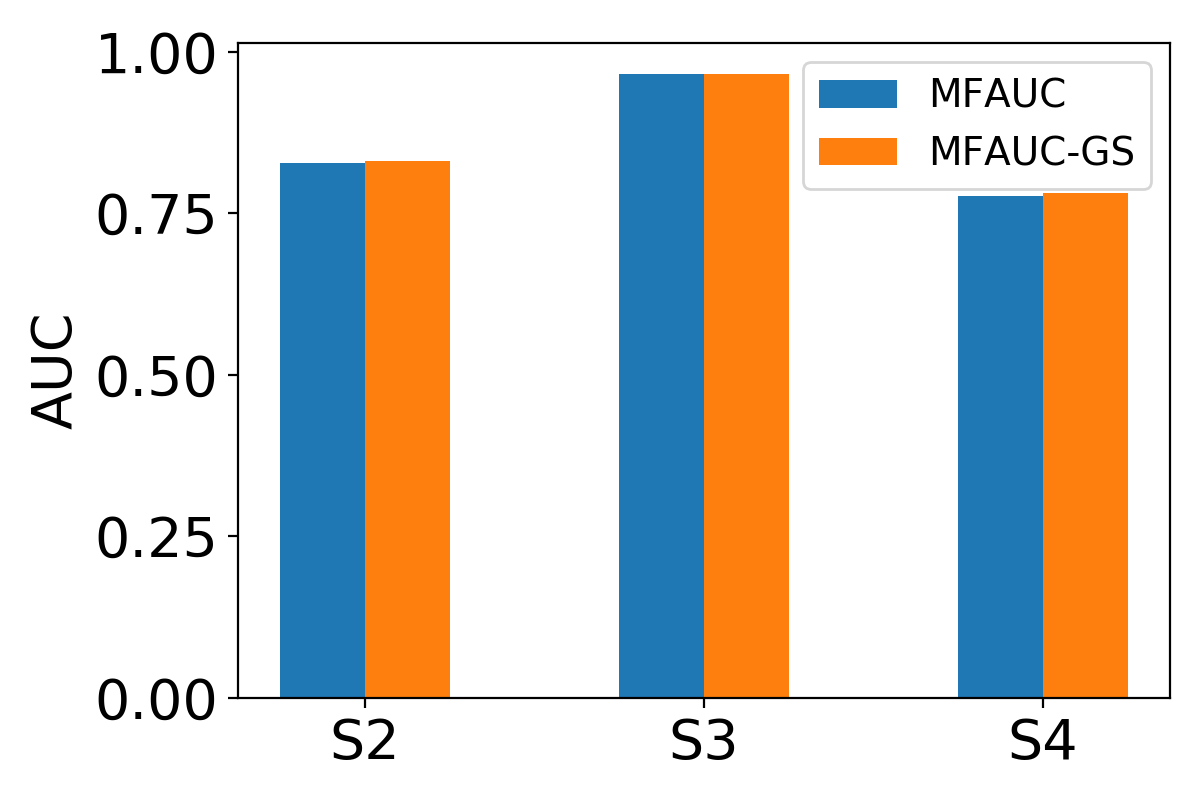}}
\hspace{1em}
\subfloat[Training time ratio (MFAUC-GS : MFAUC)]{\includegraphics[width=0.23\textwidth]{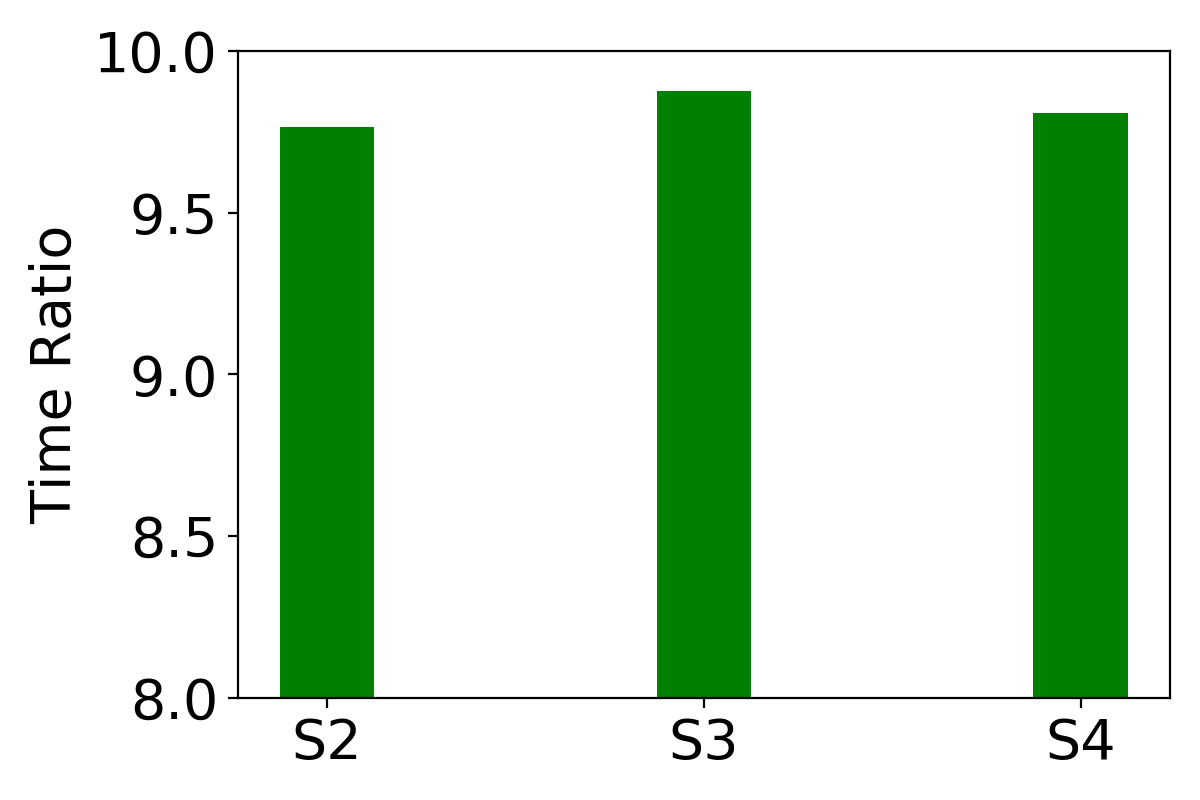}}
\caption{Comparison of MFAUC and MFAUC-GS on IC dataset, where MFAUC-GS is a variant of MFAUC that turns $\eta$ via grid search, i.e. training 10 models for 10 $\eta$ candidates respectively and choosing the one that achieves the best AUC on the test set.} 
\label{fig:Optimal_eta_MFAUC}
\end{figure}

\begin{table*}[h]
\centering
\setlength{\tabcolsep}{1pt}
\caption{Top 5 new DTIs found by MF2A from \textit{updated} golden standard datasets}
\label{tab:newDTIs2}
\begin{tabular}{@{}ccccccc@{}}
\toprule
Dataset & Drug ID & Drug name & Target ID & Target name & Score \\ \midrule
\multirow{5}{*}{NR} & D00951 & Medroxyprogesterone acetate & hsa2908 & NR3C1 & 0.674 \\
 & D00075 & Testosterone & hsa5241 & PGR & 0.672 \\
 & D00182 & Norethindrone & hsa2099 & ESR1 & 0.664 \\
 & D00040 & Cholesterol & hsa8856 & NR1I2 & 0.656 \\
 & D00075 & Testosterone & hsa2100 & ESR2 & 0.653 \\ \midrule
\multirow{5}{*}{GPCR} & D02357 & Methysergide & hsa3352 & HTR1D & 0.671 \\
 & D01713 & Epinastine hydrochloride & hsa152 & ADRA2C & 0.651 \\
 & D00426 & Risperidone & hsa150 & ADRA2A & 0.646 \\
 & D02721 & Iloprost & hsa5732 & PTGER2 & 0.641 \\
 & D01242 & Promethazine teoclate & hsa59340 & HRH4 & 0.637 \\ \midrule
\multirow{5}{*}{IC} & D00636 & Amiodarone hydrochloride & hsa3763 & KCNJ6 & 0.715 \\
 & D01450 & Bupivacaine hydrochloride & hsa3757 & KCNH2 & 0.714 \\
 & D00735 & Dyclonine hydrochloride & hsa11280 & SCN11A & 0.709 \\
 & D00349 & Isradipine & hsa773 & CACNA1A & 0.697 \\
 & D00294 & Diazoxide & hsa10060 & ABCC9 & 0.696 \\ \midrule
\multirow{5}{*}{E} & D00188 & Cholecalciferol & hsa55775 & TDP1 & 0.893 \\
 & D00043 & Isoflurophate & hsa1991 & ELANE & 0.892 \\
 & D00501 & Pentoxifylline & hsa50940 & PDE11A & 0.89 \\
 & D00364 & Loratadine & hsa7364 & UGT2B7 & 0.884 \\
 & D00437 & Nifedipine & hsa1585 & CYP11B2 & 0.883 \\  \bottomrule
\end{tabular}
\end{table*}

\bibliographystyle{IEEEtran} %
\bibliography{LB.bib}

\ifCLASSOPTIONcaptionsoff
  \newpage
\fi



%



